\documentclass[11pt]{amsart}

\usepackage[a4paper,margin=1.15in]{geometry}
\usepackage{amsmath,amssymb,amsthm,mathtools}
\usepackage[T1]{fontenc}
\usepackage{lmodern}
\usepackage{microtype}
\usepackage{enumitem}
\usepackage{booktabs,tabularx,array}
\usepackage{tikz}
\usetikzlibrary{patterns,arrows.meta,decorations.pathreplacing,calc}
\usepackage{xurl}
\usepackage[colorlinks=true,linkcolor=blue!60!black,citecolor=blue!60!black,urlcolor=blue!60!black]{hyperref}

\numberwithin{equation}{section}

\newtheorem{maintheorem}{Theorem}

\newtheorem{theorem}{Theorem}[section]
\newtheorem{proposition}[theorem]{Proposition}
\newtheorem{corollary}[theorem]{Corollary}
\newtheorem{lemma}[theorem]{Lemma}
\theoremstyle{definition}
\newtheorem{definition}[theorem]{Definition}

\theoremstyle{remark}
\newtheorem{remark}[theorem]{Remark}

\newcommand{\Z}{\mathbb Z}
\newcommand{\R}{\mathbb R}
\newcommand{\C}{\mathbb C}
\newcommand{\N}{\mathbb N}
\newcommand{\E}{\mathbb E}
\newcommand{\Sph}{\mathbb S}
\newcommand{\Bthree}{\mathsf B_3}
\newcommand{\dd}{\,\mathrm d}
\newcommand{\vt}{\vartheta}
\newcommand{\vtb}{\vartheta^{\mathrm b}}
\newcommand{\ip}[2]{\langle #1,#2\rangle}
\newcommand{\decl}[1]{{\small\nolinkurl{#1}}}
\newcommand{\Hp}{\mathbb H}
\newcommand{\one}{\mathbf 1}
\DeclareMathOperator{\sgn}{sign}

\hypersetup{pdftitle={Universal rotational invariance of two-point functions in cubic lattice models: Ising, Potts and O(N)},pdfauthor={Zhongpu Qiu},pdfkeywords={reflection positivity, rotational invariance, scaling limits, Ising model, Potts model, O(N) model}}
\begin{document}

\title[Rotational invariance of two-point functions]
{Universal rotational invariance of two-point functions in cubic lattice
models: Ising, Potts and $O(N)$}

\author{Zhongpu Qiu}
\address{Beijing Normal University, Beijing, China; and Shanghai Artificial Intelligence Laboratory, Shanghai, China}
\email{slszqzp@outlook.com}

\date{November 26, 2026}


\begin{abstract}
We prove that the two-point functions of several classical
three-dimensional lattice models become rotationally invariant on large
scales, at every temperature and in particular at the critical point. The
models include the nearest-neighbour Ising model, the $q$-state Potts and
random-cluster models with integer $q\ge2$, the XY model, and the
Heisenberg model in the state obtained by removing a uniform magnetic
field. More precisely, every subsequential scaling limit of the two-point
function, normalized by the susceptibility of a box, is invariant under
all rotations of $\R^3$. The proof rests on reflection positivity in the
coordinate and diagonal lattice planes and uses no information about
critical exponents.

By the renormalization group and an argument of Polyakov, rotational
invariance is one of the symmetries that combine into the conformal
invariance expected at critical points. Below the upper critical
dimension, rigorous results on these emergent symmetries had been
confined to two dimensions; here they are obtained in three dimensions,
for a whole class of models at once.
\end{abstract}

\maketitle

\setcounter{tocdepth}{1}
\tableofcontents

\section{Introduction}\label{sec:intro}

\subsection{Motivation}
Phase transitions are ubiquitous in physical systems. Among them,
continuous phase transitions, also called critical phenomena, are
distinguished by correlations that extend over all length scales at the
critical point. Landau's mean-field theory \cite{Landau} gave the first
general description of such transitions, and Onsager's exact solution of
the two-dimensional Ising model \cite{Onsager} showed that the actual
critical behaviour can differ from the mean-field predictions. From the
1960s on, scaling \cite{Widom,Kadanoff} and universality became the basic
concepts of the field, and the renormalization group \cite{Wilson,WilsonKogut}
provided a deep, if non-rigorous, understanding of critical behaviour.
Scaling, universality and renormalization are often called the three
pillars of the theory of critical phenomena \cite{Stanley,Fisher}.

In the renormalization-group picture, coarse-graining transformations
change the scale and the parameters of a model, and the large-scale limit
at the critical point arises as a fixed point of these transformations.
Microscopically different models may flow to the same fixed point, which
is the origin of universality. The fixed point also determines the
symmetries of the scaling limit. A lattice model has only the symmetries
of its lattice; on $\Z^3$ these are the translations by lattice vectors
and the $48$ rotations and reflections of the cube. If the critical fixed
point is unique, however, the scaling limit is invariant under all
translations, rotations and dilations of space: under dilations because
it is a fixed point, and under rotations because the lattice anisotropy
is an irrelevant perturbation, which disappears under coarse-graining.
For the three-dimensional $O(N)$ models the rate at which the cubic
anisotropy disappears has been analysed by strong-coupling,
field-theoretic and large-$N$ expansions \cite{CPRV}.

Polyakov went further and argued that a local field theory with these
symmetries is invariant under the larger group of conformal
transformations \cite{Polyakov}, namely the maps that are locally the
composition of a translation, a rotation and a dilation. In two
dimensions every holomorphic map is of this type, so that conformal
symmetry is infinite-dimensional. Belavin, Polyakov and Zamolodchikov
showed that this symmetry alone determines the critical exponents and
correlation functions of a family of two-dimensional field theories, the
minimal models, which describe for instance the critical Ising and
three-state Potts models \cite{BPZ}. In three
dimensions the conformal transformations form a finite-dimensional
group, generated by translations, rotations, dilations and the inversion
$x\mapsto x/|x|^2$. This smaller symmetry is still powerful: it is the
basis of the conformal bootstrap, which determines the critical exponents
of the three-dimensional Ising model with remarkable precision
\cite{ElShowk,PRV}. Proving rigorously that lattice models acquire these
symmetries at criticality is therefore a central problem of mathematical
statistical mechanics.

In two dimensions this problem has been solved for a small number of
models. For the critical Ising model, conformal invariance was proved
through discrete holomorphic observables \cite{Smirnov,CHI}. Such proofs
obtain conformal invariance directly, bypassing the route suggested by
field theory, which first establishes rotational and scale invariance and
then deduces conformal invariance. Along that route, rotational
invariance of the planar Ising spin--spin correlation was derived from
exact formulas \cite{Pinson}, and Duminil-Copin, Kozlowski, Krachun,
Manolescu and Oulamara proved rotational invariance of the critical
random-cluster model on the square lattice for $1\le q\le4$ by means of
the Yang--Baxter equation \cite{DKKMO}. More recently, conformal
invariance of the interfaces of the critical random-cluster model on the
square lattice, including Bernoulli bond percolation, has been announced
in \cite{OAI223}. All these arguments rely on structures that are special
to two dimensions, such as discrete holomorphicity and the Yang--Baxter
equation.

In high dimensions the picture is different. For the Ising model, as for
the $O(N)$ models with $N\ge2$, the upper critical dimension is four, and
above it the critical behaviour is expected to be described by mean-field
theory. For the Ising model and the lattice $\varphi^4$ model this has
been proved: their critical scaling limits are Gaussian in dimensions
$d>4$ \cite{Aizenman82,Frohlich82} and also in $d=4$ \cite{ADC}. In
sufficiently high dimensions the lace expansion gives the precise
asymptotics $g(x)\sim c|x|^{2-d}$ of the critical Ising two-point function
\cite{Sakai}, a power law that exhibits both the scale invariance and the
rotational invariance of its scaling limit.

Three dimensions lie between these two regimes. There is no integrable
structure, and the lace expansion does not apply. Rigorous results on the
three-dimensional critical regime are therefore scarce; for the Ising
model they include the continuity of the phase transition \cite{ADS} and
lower bounds on the critical two-point function \cite{ADC,DCP}.

In this situation the route suggested by field theory is a natural one:
establish the symmetries of the scaling limit one at a time. Translation
invariance is automatic, since the two-point function of a
translation-invariant state depends only on the difference of its
arguments. Scale invariance is a different matter: it amounts to
identifying the power law of the critical two-point function, and hence
requires control of the critical exponent. Rotational invariance asks
less. It concerns only how correlations depend on direction, not how fast
they decay with distance. It is nevertheless not automatic, because the
lattice is not rotationally invariant: a priori the two-point function
could decay differently along the coordinate axes and along the
diagonals, at every scale.

In this paper we establish this rotational invariance in three
dimensions. We prove that the two-point functions of a whole class of
three-dimensional models become rotationally invariant on large scales,
by a mechanism that requires no integrable structure, namely reflection
positivity. We first state the result for the Ising model.

\subsection{The three-dimensional Ising model}
Write $\Lambda_R=[-R,R]^3\cap\Z^3$. In the box $\Lambda_L$ give each
configuration $\sigma\in\{-1,1\}^{\Lambda_L}$ a weight proportional to
$\exp(\beta\sum\sigma_x\sigma_y)$, the sum running over nearest-neighbour
pairs in $\Lambda_L$, and let $\langle\cdot\rangle_{\beta,L}$ be the
corresponding expectation (free boundary conditions). By Griffiths'
inequalities \cite{Griffiths,KellySherman} the limit
\[
 g_\beta(x)=\lim_{L\to\infty}\langle\sigma_0\sigma_x\rangle_{\beta,L}
\]
exists; $g_\beta$ is the \emph{two-point function} of the free state. A
scaling limit of a two-point function requires a normalization. Instead of
a power of the scale, which would presuppose the critical exponent, we
normalize by the susceptibility of a box, that is, by the total
correlation between the origin and the sites of the box. For integers $R\ge1$ put
\begin{equation}\label{eq:muR}
 \chi_R=\sum_{x\in\Lambda_R}g(x),\qquad
 \mu_R=\frac1{\chi_R}\sum_{x\in\Z^3}g(x)\,\delta_{x/R},
\end{equation}
here with $g=g_\beta$. Thus $\mu_R$ is the two-point function seen at
scale $R$, as a positive measure on $\R^3$ with $\mu_R([-1,1]^3)=1$. A
\emph{tangent} of $g$ is a vague limit $\mu$ of $\mu_{R_j}$ along a
sequence of integers $R_j\to\infty$, that is,
$\int\psi\dd\mu_{R_j}\to\int\psi\dd\mu$ for every $\psi\in C_c(\R^3)$.
Tangents are the scaling limits of the two-point function, taken along
subsequences of scales.

\begin{maintheorem}[Ising model]\label{thm:ising}
Let $\beta\ge0$; in particular $\beta=\beta_c$ is allowed. Every
sequence of integers tending to infinity has a subsequence along which
$\mu_R$ converges vaguely. Every tangent $\mu$ of $g_\beta$ is invariant
under the orthogonal group $O(3)$, and it has the form
\[
 \mu=c_0\,\delta_0+f(x)\dd x,\qquad \mu([-1,1]^3)=1,
\]
where $c_0\ge0$ and $f\ge0$ is a locally integrable radial function,
continuous on $\R^3\setminus\{0\}$. Moreover, the rotational symmetry
holds at the level of lattice points: if $O\in O(3)$, $x\ne0$, and
$k_j,\ell_j\in\Z^3$ satisfy $k_j/R_j\to x$ and $\ell_j/R_j\to Ox$, then
\begin{equation}\label{eq:ising-points}
 \frac{R_j^3}{\chi_{R_j}}\bigl[g_\beta(k_j)-g_\beta(\ell_j)\bigr]\longrightarrow0 .
\end{equation}
\end{maintheorem}

The normalization in \eqref{eq:ising-points} is natural: $\chi_R/R^3$ is
the order of the largest correlation at distance $R$
(Lemma~\ref{lem:pointbound}). At $\beta_c$ the free state is the unique
Gibbs state \cite{ADS,LML}, so Theorem~\ref{thm:ising} concerns \emph{the}
critical three-dimensional Ising two-point function.

Theorem~\ref{thm:ising} determines the angular dependence of every
tangent. The radial profile $f$, and the question whether all
subsequences produce the same tangent, belong to scale invariance and
are discussed in Section~\ref{sec:remarks}. Away from the critical point
the tangents are elementary: for $\beta<\beta_c$ the two-point function is
summable \cite{ABF}, so that every tangent is the atom $\delta_0$; this is
why the atom has to be allowed in the statement. The content of
Theorem~\ref{thm:ising} lies at $\beta=\beta_c$, and since the proof uses
no information about critical exponents, it applies there without change.

\subsection{Where the rotational symmetry comes from}
Reflection positivity is one of the Osterwalder--Schrader axioms of
Euclidean field theory \cite{OS1,OS2}, and since the work of Fr\"ohlich,
Simon and Spencer and of Fr\"ohlich, Israel, Lieb and Simon it has been a
basic tool of lattice statistical mechanics
\cite{FSS,FILS1,FILS2,DLS,Biskup}. In its simplest form it says the
following. Let $\vt$ be the reflection in a plane, and let $F$ be a
function of the spins on one side of the plane. Then the correlation of
$F$ with its mirror image $\vt F$ is nonnegative,
$\langle F\cdot\vt F\rangle\ge0$. Applied to the planes orthogonal to a
fixed direction $n$, this positivity turns the translation along $n$ into
a positive self-adjoint operator, the transfer operator, and the two-point
function along $n$ acquires a spectral representation: correlations
between parallel planes at distance $t$ are Laplace transforms
$\int e^{-tE}\rho(\dd E)$ of positive measures. This is how reflection
positivity is usually used, always in one direction at a time.

The cubic lattice has two inequivalent families of planes through lattice
sites in which the nearest-neighbour Ising model is reflection positive:
the coordinate planes $\{x_i=c\}$ and the diagonal planes
$\{x_i\pm x_j=c\}$. Each family yields its own spectral representation,
along its own normals, and both describe the same two-point function. Our
main observation is that the two representations are compatible only if
the scaling limit is rotationally invariant; the mechanism is explained
in Section~\ref{sec:idea}. It uses three properties of the Ising model:
reflection positivity in these two families of planes, positivity of the
correlations, and cubic symmetry. We therefore formulate it as a theorem
about functions on $\Z^3$, and obtain Theorem~\ref{thm:ising} and its
analogues for other models as applications.

\subsection{Reflection-positive kernels}
Write $\Bthree$ for the group of the $48$ signed permutation matrices,
acting on $\Z^3$ and on $\R^3$. For $v\in\{e_1,\,e_1-e_2\}$ let $\vt_v$ be
the orthogonal reflection in the plane $v^\perp$,
\[
 \vt_{e_1}(x)=(-x_1,x_2,x_3),\qquad \vt_{e_1-e_2}(x)=(x_2,x_1,x_3),
\]
and let $\Hp_v=\{x\in\Z^3:\ x\cdot v\ge0\}$ be the closed half-space on
one side of it. We call $\vt_{e_1}$ the \emph{axial} and $\vt_{e_1-e_2}$
the \emph{diagonal site mirror}; both planes contain lattice sites. The
\emph{bond mirror} $\vtb(x)=(1-x_1,x_2,x_3)$ is the reflection in the
plane $x_1=\frac12$, which passes through midpoints of bonds.

\begin{definition}\label{def:RPkernel}
A function $g:\Z^3\to\R$ is a \emph{reflection-positive kernel} if
\begin{enumerate}[label=\textup{(K\arabic*)},leftmargin=3.2em]
\item\label{K1} $g\ge0$ and $g(0)=1$;
\item\label{K2} $g\circ P=g$ for every $P\in\Bthree$;
\item\label{K3} $g$ is positive definite: $\sum_{i,j}c_ic_j\,g(x_i-x_j)\ge0$
 for all finite families $x_i\in\Z^3$, $c_i\in\R$;
\item\label{K4} for $v=e_1$ and $v=e_1-e_2$, and all finite families
 $x_i\in\Hp_v$, $c_i\in\R$,
 \[
   \sum_{i,j}c_ic_j\,g\bigl(x_i-\vt_v x_j\bigr)\ \ge\ 0 .
 \]
\end{enumerate}
It is a \emph{strongly reflection-positive kernel} if in addition
\begin{enumerate}[label=\textup{(K\arabic*)},leftmargin=3.2em,start=5]
\item\label{K5} $\sum_{i,j}c_ic_j\,g\bigl(x_i-\vtb x_j\bigr)\ge0$
 for all finite families with $x_{i,1}\ge1$.
\end{enumerate}
\end{definition}

If $g(x-y)$ is the two-point function of a translation-invariant state,
then \ref{K3} is automatic, and \ref{K4}--\ref{K5} are exactly the
statements $\langle F\cdot\vt F\rangle\ge0$ for linear combinations $F$
of spins, for the three mirrors. The quantities $\chi_R$, $\mu_R$ and the
tangents of any such $g$ are defined by \eqref{eq:muR}.

\begin{maintheorem}[Reflection-positive kernels]\label{thm:main}
\begin{enumerate}[label=\textup{(\roman*)},leftmargin=2em]
\item Let $g$ be a reflection-positive kernel. Every sequence of integers
 tending to infinity has a subsequence along which $\mu_R$ converges
 vaguely. Every tangent $\mu$ of $g$ satisfies $\mu([-1,1]^3)=1$ and
 $O_\#\mu=\mu$ for every $O\in O(3)$. More precisely, there is a
 nonnegative $f\in L^1_{\rm loc}(\R^3)$, continuous on
 $\R^3\setminus\{0\}$, such that
 \begin{equation}\label{eq:decomp}
  \mu=c_0\,\delta_0+f(x)\dd x,\qquad c_0=\mu(\{0\}),\qquad
  f(Ox)=f(x)\quad (O\in O(3),\ x\ne0).
 \end{equation}
 Consequently $\int(\psi\circ O-\psi)\dd\mu_R\to0$ as $R\to\infty$
 through all integers, for every $\psi\in C_c(\R^3)$ and $O\in O(3)$.
\item Let $g$ be a strongly reflection-positive kernel. There is a
 constant $C$ such that $g(x)\le C\chi_r/r^3$ whenever
 $\|x\|_\infty\ge r\ge1$. Moreover, for every $O\in O(3)$, every $x\ne0$,
 every sequence of integers $R_j\to\infty$, and all $k_j,\ell_j\in\Z^3$
 with $k_j/R_j\to x$ and $\ell_j/R_j\to Ox$,
 \begin{equation}\label{eq:pointwise}
  \frac{R_j^3}{\chi_{R_j}}\bigl[g(k_j)-g(\ell_j)\bigr]\longrightarrow0.
 \end{equation}
 The same holds if $k_j/R_j$ and $\ell_j/R_j$ stay in a fixed compact
 subset of $\R^3\setminus\{0\}$ and $|k_j|/R_j-|\ell_j|/R_j\to0$.
\end{enumerate}
\end{maintheorem}

The two site mirrors carry the whole argument. The bond mirror is needed
only to compare lattice points of opposite parity, which is why it
appears in part (ii) of Theorem~\ref{thm:main} but not in part (i). The roles of the
other hypotheses are equally specific: \ref{K1} keeps all measures
positive, \ref{K2} supplies the period $\pi/2$ in the angular argument,
and \ref{K3} enters only through a bound on the mass of a box
(Lemma~\ref{lem:block}). Finally, the argument needs the \emph{harmonic gap}
of the angular argument: cubic symmetry allows in each coordinate plane only the
harmonics $e^{im\theta}$ with $4\mid m$, while the analytic continuation
grows like $e^{2|\Im\theta|}$. Since $4>2$, every nonconstant harmonic is
excluded; with a merely twofold symmetry the first allowed harmonic $m=2$
would be borderline and the argument would fail.

\subsection{Further models}
Theorem~\ref{thm:main} reduces the rotational invariance of a two-point
function to the verification of \ref{K1}--\ref{K4} (and \ref{K5} for
Theorem~\ref{thm:main}(ii)). We carry this out for the classical ferromagnetic
models. All statements hold at every $\beta\ge0$, in particular at the
critical point.

\begin{maintheorem}[Potts and random-cluster models]\label{thm:potts}
Let $q\ge2$ be an integer and $\beta\ge0$. For the nearest-neighbour
$q$-state Potts model on $\Z^3$ with bond weights
$\exp(\beta\one_{\{\sigma_x=\sigma_y\}})$, the limit
\[
 g_{q,\beta}(x-y)=\lim_{n\to\infty}
 \Bigl\langle\frac{q\one_{\{\sigma_x=\sigma_y\}}-1}{q-1}\Bigr\rangle_{D_n}
 =\phi_{p,q}(x\leftrightarrow y),\qquad p=1-e^{-\beta},
\]
exists along every increasing exhaustion of $\Z^3$ by finite sets $D_n$
with free boundary conditions, is independent of the exhaustion, and is
a strongly reflection-positive kernel; here $\phi_{p,q}$ is the free
random-cluster measure. In particular, Theorem~\ref{thm:main} holds for the connection probabilities of the free
random-cluster model with integer $q\ge2$.
\end{maintheorem}

For $q=2$ the kernel $g_{2,2\beta}$ is the Ising two-point function
$g_\beta$, so Theorem~\ref{thm:ising} is a special case of
Theorem~\ref{thm:potts}. For continuous spins let $S_x\in\Sph^{N-1}$ carry the
normalized surface measure, give each nearest-neighbour bond the weight
$\exp(\beta S_x\cdot S_y)$, and let the two-point function be
$g(x)=\E[S_0\cdot S_x]$. The internal dimension $N$ is unrelated to the
spatial dimension three on which $O(3)$ acts.

\begin{maintheorem}[Sphere spins]\label{thm:sphere}
Let $\beta\ge0$. In each of the following cases the state described is a
translation-invariant zero-field Gibbs state of the nearest-neighbour
$O(N)$ model at inverse temperature $\beta$, and its two-point function
is a strongly reflection-positive kernel; hence Theorem~\ref{thm:main} applies.
\begin{enumerate}[label=\textup{(\roman*)},leftmargin=2.2em]
\item \emph{XY model, $N=2$.} The free state, which is the common weak
 limit along every exhaustion of $\Z^3$ by finite sets with free
 boundary conditions.
\item \emph{Heisenberg model, $N=3$.} For $e\in\Sph^2$, the state
 $Q^e_\beta=\lim_{h\downarrow0}P^e_{\beta,h}$, where $P^e_{\beta,h}$ is
 the free state in the uniform field $he$. The same state is obtained
 from finite volumes with all exterior spins fixed to $e$ and a bulk
 field $h_ne$, $h_n\ge0$, $h_n\to0$, including $h_n=0$.
\item \emph{All $N\ge1$.} A Gibbs state, chosen before any scale or
 rotation, that is invariant under translations, under $\Bthree$ and
 under the internal group $O(N)$.
\item \emph{Uniqueness regime, all $N\ge1$.} The unique Gibbs state,
 whenever the Gibbs state at $\beta$ is unique (for instance if
 $6\tanh\beta<1$); it is the free state.
\end{enumerate}
\end{maintheorem}

\begin{remark}\label{rem:Heis-state}
The state in (ii) has magnetization $m_*(\beta)e$, where $m_*$ is the
spontaneous magnetization. It coincides with the free state, whose
one-point function vanishes by internal symmetry, only if $m_*(\beta)=0$;
this holds for $\beta<\beta_c$ and is open at $\beta_c$ for $N=3$. The state in
(iii) is constructed by a variational argument in
Appendix~\ref{app:selected}.
\end{remark}

What does each model have to supply? Positivity of the correlations and
reflection positivity hold in every finite box that is symmetric with
respect to a given mirror (Lemma~\ref{lem:fvRP}). The difficulty is that
different mirrors require different boxes, so one has to know that all
of them produce the \emph{same} infinite-volume two-point function. We
call this the \emph{common-limit condition}; it is formulated as
condition~(S) in Section~\ref{sec:models}. Table~\ref{tab:matrix} shows
how each hypothesis is obtained. The second column is the only place
where the specific correlation inequalities of each model (Griffiths,
Fortuin--Kasteleyn monotonicity, Ginibre, Lee--Yang,
Kunz--Pfister--Vuillermot) are used, and it determines which state
appears in Theorem~\ref{thm:sphere}.

\begin{table}[t]
\centering
\small
\renewcommand{\arraystretch}{1.15}
\begin{tabularx}{\textwidth}{@{}>{\raggedright\arraybackslash}p{2.55cm}
 >{\raggedright\arraybackslash}X
 >{\raggedright\arraybackslash}p{2.6cm}
 >{\raggedright\arraybackslash}p{2.9cm}@{}}
\toprule
Model & State; common-limit condition (S) & \ref{K1}: $g\ge0$ & \ref{K5}: weight across a bond \eqref{eq:crossing}\\
\midrule
Ising\newline (Theorem~\ref{thm:ising})
 & free state; all exhaustions agree by Griffiths' inequalities \cite{Griffiths,KellySherman}
 & Griffiths \cite{Griffiths}
 & $\cosh\beta\,(1+st\tanh\beta)$\\
\addlinespace[3pt]
Potts, random-cluster, $q\ge2$ (Theorem~\ref{thm:potts})
 & free state; all exhaustions agree by FK monotonicity (Lemma~\ref{lem:fk-mono})
 & connection probability \cite{ES}
 & $J_q+(e^\beta-1)I_q$\\
\addlinespace[3pt]
XY, $N=2$\newline (Theorem~\ref{thm:sphere}(i))
 & free state; all exhaustions agree by Ginibre's inequality (Lemma~\ref{lem:ginibre})
 & Ginibre \cite{Ginibre}
 & $\sum_\alpha\frac{\beta^{|\alpha|}}{\alpha!}s^\alpha t^\alpha$\\
\addlinespace[3pt]
Heisenberg, $N=3$ (Theorem~\ref{thm:sphere}(ii))
 & $Q^e_\beta=\lim_{h\downarrow0}P^e_{\beta,h}$; Lee--Yang \cite{FRLY} and KPV \cite{KPV} (Lemma~\ref{lem:heis-positive}, Proposition~\ref{prop:heis-ident})
 & monomial expansion (Lemma~\ref{lem:heis-positive})
 & as for XY\\
\addlinespace[3pt]
$O(N)$, $N\ge1$\newline (Theorem~\ref{thm:sphere}(iii))
 & a state selected by the variational principle \cite{Georgii} (Appendix~\ref{app:selected})
 & monomial expansion
 & as for XY, plus a Gaussian term\\
\bottomrule
\end{tabularx}
\caption{How the hypotheses of Theorem~\ref{thm:main} are
obtained. The remaining hypotheses are verified in the same way for all
models: \ref{K2} and \ref{K3} hold for the two-point function of any
translation- and $\Bthree$-invariant state, and \ref{K4} follows by
conditioning on the spins in the mirror plane (Lemma~\ref{lem:fvRP}(a)),
for any interaction. The last column writes the interaction across a bond
mirror as a positive combination of products, which is what reflection
positivity across bonds requires (Lemma~\ref{lem:fvRP}(b)); $J_q$ is the $q\times q$ matrix with all entries $1$.}
\label{tab:matrix}
\end{table}

For the Ising model, the decay bound $g(x)\le C\chi_r/r^3$ is classically
deduced from the inequalities of Messager and Miracle-Sol\'e \cite{MMS}.
In the present approach it is a consequence of reflection positivity and
positive definiteness alone (Lemma~\ref{lem:pointbound}), so the
Messager--Miracle-Sol\'e inequalities are not needed for
Theorem~\ref{thm:ising}; they are used only for the finer comparison in
Proposition~\ref{prop:ising-relative}.

\subsection{Idea of the proof}\label{sec:idea}
The proof has three steps, each of which uses one of the structures
described in the previous subsections. Let $n$ be one of the two normals
$e_1$ and $(e_1-e_2)/\sqrt2$.

\emph{Step 1: two spectral representations on the lattice.} The site
mirror orthogonal to $n$ makes the translation by one lattice step along
$n$ a symmetric contraction $T$ on a Hilbert space built from $g$. The
correlation of a packet of sites in the mirror plane with its translate
by $2j$ steps is then $\langle\varphi,T^{2j}\varphi\rangle$, the $j$th
moment of a positive measure on $[0,1]$. The two meshes differ: the
normal spacings are $2$ and $2\sqrt2$, and the cells of the mirror
lattices have areas $1$ and $\sqrt2$ (Figure~\ref{fig:meshes}). After
rescaling, every tangent inherits a positive-time spectral
representation $K_\varphi(t)=\int e^{-tE}\rho_\varphi(\dd E)$ in each of
the two directions.

\emph{Step 2: Stieltjes functions on lines.} In Fourier variables the
spectral representation of Step~1 says the following. If $C$ is the
Fourier transform of the tangent and $k\ne0$ is a frequency orthogonal to
$n$, then $\omega\mapsto C(\omega n+k)$ is a Stieltjes function of
$\omega^2$,
\[
 C(\omega n+k)=b+\int\frac{\xi(\dd s)}{\omega^2+s},\qquad b\ge0,\ \xi\ge0 .
\]
Such a function extends holomorphically to $\C\setminus(-\infty,0]$, with
bounds that depend only on its value at a single point.

\begin{figure}[t]
\centering
\begin{tikzpicture}[x=1.45cm,y=1.15cm,>=Stealth]
 \def\Y{2.15}
 \foreach \a/\b in {0.5/1.5,2.5/3.5}{
   \fill[blue!10] (\a,-\Y) rectangle (\b,\Y);}
 \foreach \a/\b in {-0.5/0.5,1.5/2.5,3.5/4.5}{
   \fill[orange!14] (\a,-\Y) rectangle (\b,\Y);}
 \foreach \x in {0,2,4}{
   \draw[blue!70!black,dashed,thick] (\x,-\Y) -- (\x,\Y);}
 \foreach \x in {1,3}{
   \draw[orange!80!black,densely dotted,very thick] (\x,-\Y) -- (\x,\Y);}
 \draw[->] (-0.65,0) -- (4.75,0) node[right] {$\Re\theta$};
 \draw[->] (-0.5,-\Y-0.15) -- (-0.5,\Y+0.25) node[left] {$\Im\theta$};
 \foreach \x/\l in {0/0,1/\frac\pi4,2/\frac\pi2,3/\frac{3\pi}4,4/\pi}{
   \draw (\x,0.06) -- (\x,-0.06);
   \node[above=1pt,font=\scriptsize] at (\x,\Y) {$\l$};}
 \draw[very thick,->,black!80] (0,0) -- (2,0);
 \draw[very thick,->,black!80] (2,0) -- (2,-1.55);
 \draw[very thick,->,black!80] (2,-1.55) -- (0,-1.55);
 \draw[very thick,->,black!80] (0,-1.55) -- (0,0);
 \node[fill=white,inner sep=1.5pt,font=\scriptsize] at (1,-1.82)
   {$\Im\theta=-Y$:\ $|\Psi\,e^{-im\theta}|\le Ce^{(2-|m|)Y}$};
 \node[font=\scriptsize,blue!60!black,fill=white,inner sep=1pt,align=center] at (1,1.05)
   {axial\\ chart};
 \node[font=\scriptsize,orange!60!black,fill=white,inner sep=1pt,align=center] at (2,1.05)
   {diagonal\\ chart};
 \node[font=\scriptsize,blue!60!black,fill=white,inner sep=1pt,align=center] at (3,1.05)
   {axial\\ chart};
 \draw[decorate,decoration={brace,amplitude=4pt,mirror}] (4.62,0.08) -- (4.62,\Y)
   node[midway,right=5pt,font=\scriptsize,align=left] {$|\Psi(\theta)|\le$\\ $Ce^{2|\Im\theta|}$};
\end{tikzpicture}
\caption{The complex $\theta$-plane in Step~3. Dashed lines: the
boundaries $\Re\theta\in\frac\pi2\Z$ of the charts attached to the axial
normal $e_1$. Dotted lines: those of the charts attached to the diagonal
normal, $\Re\theta\in\frac\pi4+\frac\pi2\Z$. Each chart is holomorphic on
the open strip between two consecutive lines of its own family; since the
two families interlace, the charts overlap on vertical strips and glue to
an entire function. Shading indicates, for each value of $\Re\theta$, the
chart used for the growth bound, namely the one with
$|\sin2(\theta-\alpha)|\ge1/\sqrt2$; these bands keep a distance $\pi/8$
from the boundaries of their chart. The rectangle is the contour used to
eliminate the harmonic $e^{im\theta}$, $m\in4\Z\setminus\{0\}$.}
\label{fig:strips}
\end{figure}
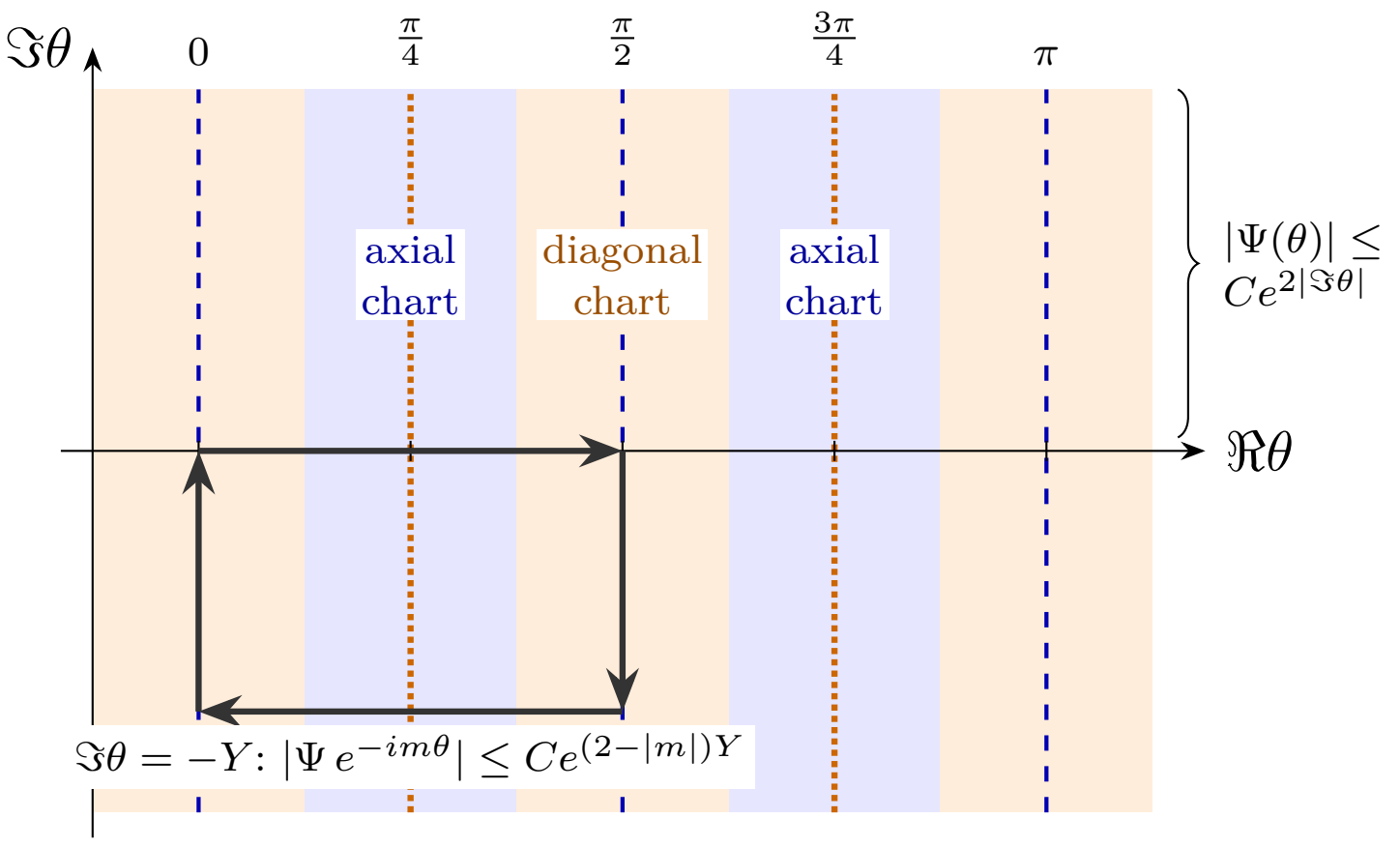

\emph{Step 3: an entire function of the angle.} Fix a frequency
$z_0\ne0$ along $e_3$ and consider
$C(e^v\cos\theta,e^v\sin\theta,z_0)$, averaged over $v=\log|p|$ against
a Gaussian. In polar coordinates adapted to $n$, the variable of the
Stieltjes function of Step~2 is $\cot^2(\theta-\alpha)$, where
$\alpha\in\{0,-\pi/4\}$ is the angle of $n$. This continues
holomorphically to each vertical strip between consecutive lines
$\Re\theta\in\alpha+\frac\pi2\Z$. The lines for $\alpha=0$ and those for
$\alpha=-\pi/4$ interlace, so the two families of charts glue to an
entire function $\Psi(\theta)$ of period $\pi/2$ (Figure~\ref{fig:strips}).
The bound of Step~2 gives $|\Psi(\theta)|\le Ce^{2|\Im\theta|}$, and moving
the contour in the Fourier coefficient of order $m\in4\Z\setminus\{0\}$
produces a factor $e^{(2-|m|)Y}$, which tends to zero. Hence $\Psi$ is
constant. Removing the Gaussian average gives invariance under rotations
in each coordinate plane, hence under $O(3)$.

Two points require care. First, the tangents are only known to exist
along subsequences, and the spectral measures may depend on the
subsequence and on the packet; the proof works with a single fixed
tangent throughout. Second, the two-point function may oscillate on the
lattice scale between the two parity classes along a coordinate axis, a
possibility that the site mirrors alone do not exclude. For the measure
statement of Theorem~\ref{thm:main}(i) this is removed by an averaging that does
not change the tangents (Section~\ref{sec:site}); for the lattice-point
statement of Theorem~\ref{thm:main}(ii) we use the bond mirror.

\subsection{Relation to previous work}
In Euclidean field theory, reflection positivity yields, through the
Osterwalder--Schrader reconstruction \cite{OS1,OS2,GlimmJaffe}, a positive
transfer operator and a spectral representation in the time direction,
while Euclidean invariance is imposed as a separate axiom. On the lattice,
infrared bounds and chessboard estimates \cite{FSS,FILS1,FILS2,DLS,Biskup}
use reflections in one family of planes at a time in order to prove
phase transitions. Here reflection positivity in two inequivalent
families of planes, together with the cubic symmetry that relates them,
produces the rotational symmetry that the lattice lacks.

For the Ising model, the inequalities of Messager and Miracle-Sol\'e
\cite{MMS} compare the two-point function in different lattice directions
(see also \cite[\S4.1]{ADC}); Theorem~\ref{thm:ising} identifies the
angular dependence of its scaling limit. Lower bounds on the critical two-point
function \cite{ADC,DCP} are used only in
Proposition~\ref{prop:ising-relative}. In two dimensions the works
\cite{Pinson,Smirnov,CHI,DKKMO} obtain stronger conclusions, namely
explicit correlation functions, conformal invariance, or rotational
invariance of all large-scale observables, using integrable structure;
see also \cite{ADTW} for emergent planarity in finite-range
two-dimensional Ising models. The present method is restricted to two-point
functions, but it requires no integrable structure and applies in three
dimensions to several models at once.

\subsection{Organization of the paper}
Sections~\ref{sec:lattice}--\ref{sec:site} prove Theorem~\ref{thm:main}. Section~\ref{sec:lattice} derives the lattice
consequences of \ref{K1}--\ref{K5}: the transfer operators, the moment
representations, difference bounds and the decay bound; these are
collected in the notion of an \emph{admissible kernel}
(Definition~\ref{def:admissible}). Section~\ref{sec:limits} passes from an
admissible kernel to its tangents and constructs their positive-time
spectral representations and Fourier transforms.
Section~\ref{sec:angular} contains the angular argument (Steps~2 and~3
above) and the proof of Theorem~\ref{thm:main}(ii). Section~\ref{sec:site} removes
the bond mirror and proves Theorem~\ref{thm:main}(i). Section~\ref{sec:models}
verifies the hypotheses for the models of Theorems~\ref{thm:ising},
\ref{thm:potts} and~\ref{thm:sphere}, and Section~\ref{sec:remarks} discusses open
problems. Appendix~\ref{app:selected} constructs the state of
Theorem~\ref{thm:sphere}(iii), and Appendix~\ref{app:freecube} treats a
prescribed free-cube limit for sphere spins.

\section{Lattice consequences of reflection positivity}\label{sec:lattice}

This section contains everything that is proved on the lattice itself.
Starting from \ref{K1}--\ref{K5}, we construct the transfer operators
attached to the site mirrors (Lemma~\ref{lem:T}), deduce the moment
representations of Step~1 of Section~\ref{sec:idea}
(Corollary~\ref{cor:moments}), and derive from them bounds on lattice
differences and on the decay of the kernel. The output is the notion of an
\emph{admissible kernel} (Definition~\ref{def:admissible}), which is all
that the later sections use.

\subsection{Notation}
We write $\|x\|_\infty=\max_i|x_i|$ and $|x|$ for the Euclidean norm.
For $g:\Z^3\to\R$ and real $r>0$ put
\[
 U_g(r)=\sup_{\|x\|_\infty\ge r}g(x),\qquad
 \chi^g_R=\sum_{x\in\Lambda_R}g(x),
\]
and drop the index $g$ when no confusion can arise. For $v\in\Z^3$,
$\Delta_vg(x)=g(x+v)-g(x)$, and $\Delta^j_v$ is its $j$th iterate. Let
\[
 \mathcal D=\Bthree e_1\ \cup\ \Bthree(e_1-e_2)
\]
be the six axial and twelve diagonal lattice vectors. For
$v\in\mathcal D$, $\vt_v$ denotes the reflection in $v^\perp$ and
$\Hp_v=\{x\in\Z^3:x\cdot v\ge0\}$. If $v=Pv_0$ with $P\in\Bthree$ and
$v_0\in\{e_1,e_1-e_2\}$, then $\vt_v=P\vt_{v_0}P^{-1}$ and
$\Hp_v=P\Hp_{v_0}$. Hence, by \ref{K2}, the inequality in \ref{K4} holds
for every $v\in\mathcal D$, and \ref{K5} holds for the images under
$\Bthree$ of the bond mirror. Since only differences of sites enter,
the same inequalities hold for all translates of these mirrors by
lattice vectors.

\subsection{Transfer operators}
Fix a reflection-positive kernel $g$ and $v\in\mathcal D$. On the space
$V_v$ of finitely supported functions $a:\Hp_v\to\R$ put
\[
 \ip ab_v=\sum_{x,y\in\Hp_v}a(x)\,b(y)\,g(x-\vt_vy),
\qquad (T_va)(x)=a(x-v)\ \text{for }x\in\Hp_v .
\]
Writing $[x]$ for the indicator of $\{x\}$, $T_v[x]=[x+v]$; note that
$\Hp_v+v\subset\Hp_v$. By \ref{K4} the form $\ip\cdot\cdot_v$ is positive
semidefinite. We call $T_v$ the \emph{transfer operator} in the direction
$v$.

\begin{lemma}\label{lem:T}
\begin{enumerate}[label=\textup{(\roman*)},leftmargin=2em]
\item $T_v$ is symmetric and $\ip{T_va}{T_va}_v\le\ip aa_v$. Hence $T_v$
 induces a self-adjoint contraction, still denoted $T_v$, on the Hilbert
 space $\mathcal H_v$ obtained from $V_v$ by quotienting null vectors and
 completing; $\sigma(T_v)\subset[-1,1]$.
\item If $v$ is diagonal, then $\ip a{T_va}_v\ge0$, so
 $\sigma(T_v)\subset[0,1]$.
\item If $v$ is axial and $g$ satisfies \ref{K5}, then
 $\sigma(T_v)\subset[0,1]$.
\end{enumerate}
\end{lemma}

\begin{proof}
Since $\vt_vv=-v$, $\ip{[x]}{T_v[y]}_v=g(x-\vt_vy+v)=\ip{T_v[x]}{[y]}_v$.
Let $m_j=\ip{T_v^ja}{T_v^ja}_v$. By $|g|\le1$, $0\le m_j\le(\sum|a(x)|)^2$,
and by symmetry and Cauchy--Schwarz $m_j^2=\ip{T^{j-1}_va}{T^{j+1}_va}_v^2\le
m_{j-1}m_{j+1}$. If $m_0=0$ then $m_1^2\le m_0m_2=0$. If $m_1>m_0>0$, the
ratios $m_{j+1}/m_j$ are nondecreasing and the sequence grows at least
geometrically, which contradicts boundedness. Hence $m_1\le m_0$. In
particular $T_v$ maps null vectors to null vectors, and (i) follows.

For (ii) it suffices by \ref{K2} to take $v=e_1-e_2$. Then
$e_1-\vt_ve_1=v$, so for $x,y\in\Hp_v$,
\[
 x-\vt_vy+v=(x+e_1)-\vt_v(y+e_1),\qquad (x+e_1)\cdot v=x\cdot v+1\ge0 .
\]
Thus $\ip a{T_va}_v$ is the form of \ref{K4} for the translated family
$x+e_1$, and is nonnegative. A symmetric contraction with
$\ip a{Ta}\ge0$ on a dense set has spectrum in $[0,1]$.

For (iii) take $v=e_1$. For $x,y\in\Hp_{e_1}$ put $x'=x+e_1$, $y'=y+e_1$,
so $x'_1,y'_1\ge1$, and
$x-\vt_{e_1}y+e_1=(x_1+y_1+1,x_2-y_2,x_3-y_3)=x'-\vtb y'$. Thus
$\ip a{T_{e_1}a}_{e_1}$ is the form of \ref{K5}.
\end{proof}

\begin{corollary}[Moment representation]\label{cor:moments}
Let $v\in\mathcal D$ and let $a$ be a finitely supported real function on
$\Hp_v$. Then
\[
 s_n=\sum_{x,y}a(x)a(y)\,g(x-\vt_vy+2nv)=\ip a{T_v^{2n}a}_v,\qquad n\ge0,
\]
is the moment sequence $s_n=\int_{[0,1]}\lambda^n\,\nu_a(\dd\lambda)$ of a
positive measure $\nu_a$ of total mass $s_0\le(\sum|a(x)|)^2$. If
$\sigma(T_v)\subset[0,1]$, the same holds for $n\mapsto\ip a{T_v^na}_v$.
\end{corollary}

\begin{proof}
$\nu_a$ is the spectral measure of $T_v^2\ge0$ at $a$; the identity
$\ip a{T^{2n}_va}_v=\sum a(x)a(y)g(x-\vt_v(y+2nv))$ and $\vt_vv=-v$ give the
formula.
\end{proof}

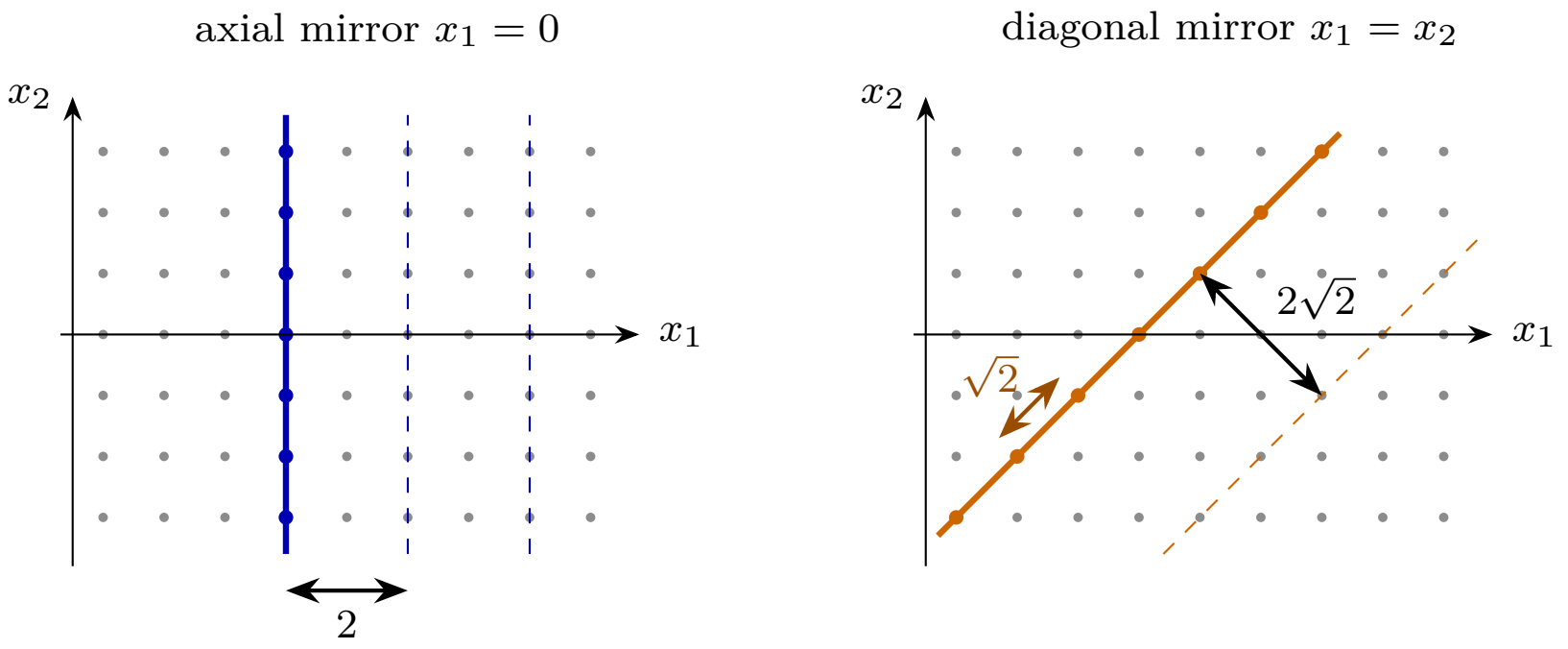
\begin{figure}[t]
\centering
\begin{tikzpicture}[scale=0.42,>=Stealth]
 \begin{scope}
  \foreach \i in {-3,...,5}{\foreach \j in {-3,...,3}{
     \fill[black!45] (\i,\j) circle (2.2pt);}}
  \draw[blue!70!black,very thick] (0,-3.6) -- (0,3.6);
  \foreach \i in {2,4}{\draw[blue!70!black,dashed] (\i,-3.6) -- (\i,3.6);}
  \foreach \j in {-3,...,3}{\fill[blue!70!black] (0,\j) circle (3.4pt);}
  \draw[<->,thick] (0,-4.2) -- node[below] {\scriptsize $2$} (2,-4.2);
  \node[font=\scriptsize,align=center] at (1.5,5.0) {axial mirror $x_1=0$};
  \draw[->] (-3.7,0) -- (5.8,0) node[right] {\scriptsize$x_1$};
  \draw[->] (-3.5,-3.8) -- (-3.5,3.9) node[left] {\scriptsize$x_2$};
 \end{scope}
 \begin{scope}[xshift=14cm]
  \foreach \i in {-3,...,5}{\foreach \j in {-3,...,3}{
     \fill[black!45] (\i,\j) circle (2.2pt);}}
  \draw[orange!80!black,very thick] (-3.3,-3.3) -- (3.3,3.3);
  \draw[orange!80!black,dashed] (0.4,-3.6) -- (5.6,1.6);
  \foreach \u in {-3,...,3}{\fill[orange!80!black] (\u,\u) circle (3.4pt);}
  \draw[<->,thick] (1,1) -- node[above right=-1pt] {\scriptsize $2\sqrt2$} (3,-1);
  \draw[<->,thick,orange!60!black] (-2.3,-1.7) -- node[above left=-2pt] {\scriptsize $\sqrt2$} (-1.3,-0.7);
  \node[font=\scriptsize,align=center] at (1.5,5.0) {diagonal mirror $x_1=x_2$};
  \draw[->] (-3.7,0) -- (5.8,0) node[right] {\scriptsize$x_1$};
  \draw[->] (-3.5,-3.8) -- (-3.5,3.9) node[left] {\scriptsize$x_2$};
 \end{scope}
\end{tikzpicture}
\caption{The two site mirrors in the plane $x_3=0$. Large dots: the
mirror lattice $\Gamma_v$ (it extends in the $x_3$ direction with unit
spacing, so its cell area is $1$, respectively $\sqrt2$). Dashed: the
sampled layers $\Gamma_v+2jv$ ($j=1,2$ on the left, $j=1$ on the right),
at normal spacing $2$, respectively $2\sqrt2$.}
\label{fig:meshes}
\end{figure}

\begin{remark}[The two meshes]\label{rem:meshes}
For $a$ supported on the mirror, $x-\vt_vy=x-y$. For $v=e_1$ the mirror
lattice is $\Gamma_{e_1}=\{(0,u,w)\}$, of cell area $1$, and the normal
translation $2e_1$ has length $2$. For $v=e_1-e_2$ the mirror lattice is
$\Gamma_{e_1-e_2}=\{(u,u,w)\}$, of cell area $\sqrt2$, and the sampled
displacements are $(u+2n,u-2n,w)$: the normal translation $2(e_1-e_2)$
has length $2\sqrt2$ (Figure~\ref{fig:meshes}). Both numbers enter the
continuum normalization in Section~\ref{sec:spectral}. The signed
permutations give three axial and six diagonal unoriented normals.
\end{remark}

\subsection{Difference bounds}
\begin{lemma}\label{lem:differences}
Let $g$ be a reflection-positive kernel, $v\in\mathcal D$, $x\in\Hp_v$ and
$k=\lfloor x\cdot v/|v|^2\rfloor\ge6$. For integers $l,j\ge1$,
\begin{equation}\label{eq:diff}
 \bigl|\Delta^j_{lv}g(x)\bigr|\le (3j)^j\Bigl(\frac lk\Bigr)^jU_g\bigl(2\lfloor k/3\rfloor\bigr),
\end{equation}
provided $l$ is even or $\sigma(T_v)\subset[0,1]$.
\end{lemma}

\begin{proof}
Put $y=x-kv$, so that $0\le y\cdot v<|v|^2$ and $y\in\Hp_v$. Since
$\ip{[y]}{T_v^n[0]}_v=g(y+nv)$ for $n\ge0$,
\[
 \Delta^j_{lv}g(x)=\ip{[y]}{T_v^k(T_v^l-1)^j[0]}_v
 =\ip{T_v^a[y]}{T_v^{k-2a}(T_v^l-1)^jT_v^a[0]}_v,\qquad a=\lfloor k/3\rfloor .
\]
Now $\|T^a_v[y]\|^2=g(y-\vt_vy+2av)$ and $y-\vt_vy=cv$ with $c\in[0,2)$;
since $\|v\|_\infty=1$, this displacement has sup-norm at least $2a$.
Likewise $\|T^a_v[0]\|^2=g(2av)$. Both are at most $U_g(2a)$. Under
either proviso, for $t\in\sigma(T_v)$ we have $|1-t^l|=1-|t|^l\le
l(1-|t|)$, and with $n=k-2a\ge k/3$,
\[
 \sup_{0\le s\le1}s^n\,l^j(1-s)^j\le l^j\Bigl(\frac j{n+j}\Bigr)^j\le
 \Bigl(\frac{3jl}{k}\Bigr)^j. \qedhere
\]
\end{proof}

\subsection{A block bound}
The following is the only place where \ref{K3} is used.

\begin{lemma}\label{lem:block}
Let $h:\Z^3\to[0,\infty)$ be even and positive definite. For every
integer $m\ge1$ and $b\in\Z^3$,
\[
 \sum_{\|x-b\|_\infty\le m/2}h(x)\ \le\ 8\,\chi^h_m .
\]
Consequently $\chi^h_{AR}\le C_A\chi^h_R$ for every $A\ge1$, and
$\sup_R\frac1{\chi^h_R}\sum_{\|x\|_\infty\le LR}h(x)<\infty$ for each $L$.
\end{lemma}

\begin{proof}
Let $B=[0,m)^3\cap\Z^3$. Positive definiteness applied to
$\one_B-\one_{B+b}$ and evenness give
\[
 \sum_{u\in B,\ u'\in B+b}h(u'-u)\le\sum_{u,u'\in B}h(u-u')\le m^3\chi^h_m .
\]
The left side equals $\sum_xh(x)\prod_{i=1}^3(m-|x_i-b_i|)_+$, and the
weight is at least $(m/2)^3$ when $\|x-b\|_\infty\le m/2$. The
consequences follow by covering $\Lambda_{AR}$ by at most
$(2A+1)^3$ cubes of the form $\{\|x-b\|_\infty\le R/2\}$.
\end{proof}

\subsection{Admissible kernels}
The proof of the angular theorem uses only the following properties.

\begin{definition}\label{def:admissible}
A function $h:\Z^3\to[0,1]$ with $h(0)>0$, invariant under $\Bthree$ and
positive definite, is \emph{admissible with delay $d\in\N_0$} if there
are constants $C_0\ge1$ and $r_0\ge1$ such that, writing $U=U_h$:
\begin{enumerate}[label=\textup{(O\arabic*)},leftmargin=3em]
\item\label{O1} for real $r\ge r_0$, all $x$ with $\|x\|_\infty\ge r$ and
 all $z$ with $\|z\|_\infty\le r/C_0$,
 \[|h(x+z)-h(x)|\le C_0\,\frac{\|z\|_\infty}r\,U(r/C_0);\]
\item\label{O2} for real $r\ge r_0$, all $v\in\mathcal D$, all $x$ with
 $x\cdot v\ge r$, all $1\le j\le4$ and all integers $1\le l\le r/C_0$,
 \[|\Delta^j_{2lv}h(x)|\le C_0\Bigl(\frac lr\Bigr)^jU(r/C_0);\]
\end{enumerate}
\begin{enumerate}[label=\textup{(M$_d$)},leftmargin=3em]
\item\label{Md} for $v\in\{e_1,e_1-e_2\}$ and every finitely supported real
 $a$ on the mirror lattice $\Gamma_v=v^\perp\cap\Z^3$, the sequence
 $n\mapsto\sum_{x,y}a(x)a(y)\,h\bigl(x-y+2(n+d)v\bigr)$, $n\ge0$, is the
 moment sequence of a finite positive measure on $[0,1]$.
\end{enumerate}
\end{definition}

\begin{proposition}\label{prop:admissible}
Every reflection-positive kernel satisfies \ref{O2} and
\textup{(M$_0$)}. Every strongly reflection-positive kernel is
admissible with delay $0$.
\end{proposition}

\begin{proof}
(M$_0$) is Corollary~\ref{cor:moments}. For \ref{O2}, if $x\cdot v\ge
r\ge36$ then $k=\lfloor x\cdot v/|v|^2\rfloor\ge r/3\ge6$ and
$2\lfloor k/3\rfloor\ge r/6$; apply \eqref{eq:diff} with the even step $2l$.

For \ref{O1} assume \ref{K5}. By Lemma~\ref{lem:T}(ii),(iii) and \ref{K2},
$\sigma(T_v)\subset[0,1]$ for all $v\in\mathcal D$. Let $\|x\|_\infty\ge
r$ and $\|z\|_\infty\le r/40$. Applying an element of $\Bthree$ to $x$
and $z$ simultaneously, we may assume $x_1=\|x\|_\infty$ and
$x_2,x_3\ge0$. Write $z=a_1e_1+a_2(e_1+e_2)+a_3(e_1+e_3)$ with
$a_2=z_2$, $a_3=z_3$, $a_1=z_1-z_2-z_3$, so that
$|a_1|+|a_2|+|a_3|\le5\|z\|_\infty\le r/8$. Walk from $x$ to $x+z$ by
unit steps, first along $\pm e_1$, then $\pm(e_1+e_2)$, then
$\pm(e_1+e_3)$. Every point $w$ visited satisfies $\|w-x\|_\infty\le r/8$,
hence $w\cdot e_1\ge7r/8$ and $w\cdot(e_1+e_i)\ge3r/4$ for $i=2,3$. Bound
each step $w\to w\pm v$ by \eqref{eq:diff} with $j=l=1$, applied at the
endpoint from which the step points in the direction $+v$. For $r\ge48$
this gives at most $(C/r)\,U_g(r/9)$ per step, and summing over at most
$5\|z\|_\infty$ steps gives \ref{O1}.
\end{proof}

\subsection{The pointwise bound}
\begin{lemma}\label{lem:pointbound}
Let $h$ be admissible. There is $C$ such that for all real $r\ge1$
\[
 U_h(r)\le C\,\chi^h_{\lfloor r\rfloor}/r^3 .
\]
In particular, for $0<\alpha\le1$ and integers $R\ge1$,
$\sup_{\|x\|_\infty\ge\alpha R}h(x)\le C\alpha^{-3}\chi^h_R/R^3$.
\end{lemma}

\begin{proof}
Put $V(r)=r^3U(r)/\chi_{\lfloor r\rfloor}$, $D=C_0$ and $K=2D^3$, and let
$\delta=1/(4C_0K)$. Suppose $r\ge r_0D$, $\delta r\ge1$, $U(r)>0$ and
$U(r/D)\le KU(r)$. Pick $x$ with $\|x\|_\infty\ge r$ and $h(x)\ge\frac34U(r)$.
By \ref{O1}, for $\|z\|_\infty\le\delta r$,
$|h(x+z)-h(x)|\le C_0\delta\,U(r/D)\le\frac14U(r)$, so $h\ge\frac12U(r)$
on at least $(\delta r)^3$ sites within distance $\lfloor r\rfloor/2$ of
$x$. Lemma~\ref{lem:block} gives $V(r)\le16\delta^{-3}$. If instead
$U(r/D)>KU(r)$, then, $\chi$ being nondecreasing,
$V(r)\le(D^3/K)V(r/D)=\frac12V(r/D)$. Iterating, either the first case
occurs at some $r/D^i$, or $r/D^i$ falls below
$r_1=\max(r_0D,\delta^{-1})$, where $V\le r_1^3/h(0)$ because $h\le1$
and $\chi\ge h(0)$. Hence $V(r)\le\max(16\delta^{-3},r_1^3/h(0))$. The
last assertion follows from monotonicity of $\chi$.
\end{proof}

\section{Tangents of admissible kernels}\label{sec:limits}

We now pass from the lattice to the continuum. For an admissible kernel
we construct its tangents together with a continuous density away from
the origin (Proposition~\ref{prop:spatial}), transfer the moment
representations to a positive-time spectral representation of each
tangent (Proposition~\ref{prop:spectral}), and show that each tangent has
a Fourier transform that is a continuous function away from the origin
(Proposition~\ref{prop:fourier}). These are the three inputs of the
angular argument in Section~\ref{sec:angular}.

Throughout this section $h$ is admissible with delay $d$, and $\chi_R$,
$\mu_R$, $U$ refer to $h$. Put
\[
 F_R(x)=\frac{R^3}{\chi_R}\,h(x)\qquad(x\in\Z^3),
\]
so that $\mu_R=R^{-3}\sum_xF_R(x)\,\delta_{x/R}$.

\subsection{Compactness and the off-origin density}\label{sec:spatial}
\begin{proposition}\label{prop:spatial}
Every sequence of integers $R_j\to\infty$ has a subsequence, still
denoted $R_j$, and a continuous $f:\R^3\setminus\{0\}\to[0,\infty)$ with
the following properties.
\begin{enumerate}[label=\textup{(\alph*)},leftmargin=2em]
\item For every compact $K\subset\R^3\setminus\{0\}$,
 $\sup\{|F_{R_j}(x)-f(x/R_j)|:\ x\in\Z^3,\ x/R_j\in K\}\to0$.
\item $\mu_{R_j}\to\mu=c_0\delta_0+f\dd x$ vaguely, where $c_0\ge0$,
 $f\in L^1_{\rm loc}(\R^3)$ and $\mu([-1,1]^3)=1$. If a vague limit of
 $\mu_{R_j}$ is prescribed, it is this $\mu$.
\item $f\circ P=f$ for $P\in\Bthree$; $f(x)\le C$ for $\|x\|_\infty\ge1$;
 and for $L\ge1$, $v\in\mathcal D$, $x\cdot v\ge L$, $1\le j\le4$ and
 $0<t\le L/C$,
 \[
   \bigl|\Delta^j_{tv}f(x)\bigr|\le C\,(t/L)^j,\qquad
   \Delta_{tv}f(x)=f(x+tv)-f(x).
 \]
\end{enumerate}
\end{proposition}

\begin{proof}
Fix $\alpha\in(0,1)$ and consider sites with
$\alpha R\le\|x\|_\infty\le R/\alpha$. By Lemma~\ref{lem:pointbound},
$F_R(x)\le C\alpha^{-3}$ there. By \ref{O1} with $r=\alpha R$ and
Lemma~\ref{lem:pointbound} applied at radius $\alpha R/C_0$,
\[
 |F_R(x+z)-F_R(x)|\le C_0\frac{\|z\|_\infty}{\alpha R}\cdot
 \frac{R^3U(\alpha R/C_0)}{\chi_R}\le C_\alpha\frac{\|z\|_\infty}R
 \qquad(\|z\|_\infty\le\alpha R/C_0).
\]
Extend $x/R\mapsto F_R(x)$ from $R^{-1}\Z^3$ to $\R^3$ by linear
interpolation on a fixed simplicial subdivision of the cubes of
$R^{-1}\Z^3$. The interpolants are uniformly bounded and uniformly
Lipschitz on every compact subset of $\R^3\setminus\{0\}$. By the
Arzel\`a--Ascoli theorem and a diagonal argument over $\alpha=2^{-m}$ we
obtain a subsequence and a continuous limit $f\ge0$ with (a).

By Lemma~\ref{lem:block}, $\mu_R([-L,L]^3)\le C_L$ uniformly in $R$, so
after a further extraction $\mu_{R_j}$ converges vaguely to a Radon
measure $\mu$. If $\psi\in C_c(\R^3\setminus\{0\})$, then
$\int\psi\dd\mu_R=R^{-3}\sum_xF_R(x)\psi(x/R)\to\int\psi f\dd x$ by (a)
and Riemann sums. Hence $\mu=f\dd x$ on $\R^3\setminus\{0\}$, and $f$ is
locally integrable because $\mu$ is Radon. Put $c_0=\mu(\{0\})$. The
boundary of $[-1,1]^3$ is compact, has Lebesgue measure zero and avoids
the origin, so it is $\mu$-null, and
$\mu([-1,1]^3)=\lim\mu_{R_j}([-1,1]^3)=1$. If a vague limit was
prescribed, we started from the sequence realizing it.

For (c), symmetry is inherited from $h$, and the bound on $f$ follows
from Lemma~\ref{lem:pointbound} with $\alpha=\frac12$. For the
differences take $x_j\in\Z^3$ with $x_j/R_j\to x$, and let $l_j$ be the
nearest positive integer to $tR_j/2$. With $r=LR_j/2$ we have
$x_j\cdot v\ge r$ and $l_j\le r/C_0$ for large $j$ if $t\le L/(2C_0)$,
so by \ref{O2} and Lemma~\ref{lem:pointbound}
\[
 \bigl|\Delta^j_{2l_jv}F_{R_j}(x_j)\bigr|
 \le C_0\Bigl(\frac{l_j}r\Bigr)^j\frac{R_j^3\,U(R_j/(2C_0))}{\chi_{R_j}}
 \le C\Bigl(\frac tL\Bigr)^j .
\]
The points $(x_j+2il_jv)/R_j$, $0\le i\le4$, converge to $x+itv$, which
satisfy $(x+itv)\cdot v\ge L$ and so lie in a fixed compact set avoiding
the origin. By (a) the left side converges to $|\Delta^j_{tv}f(x)|$.
\end{proof}

We fix such a tangent $\mu$, its density $f$, and the scale sequence for
the rest of Sections~\ref{sec:limits} and~\ref{sec:angular}.

\subsection{Positive-time spectral representation}\label{sec:spectral}
Let $v\in\{e_1,e_1-e_2\}$, $n=v/|v|$, and identify $n^\perp$ with $\R^2$
through orthonormal coordinates. For $\varphi\in C^\infty_c(n^\perp;\C)$
put
\[
 A_\varphi(u)=\int_{n^\perp}\overline{\varphi(w)}\,\varphi(w+u)\dd w,
 \qquad
 K_\varphi(t)=\int_{n^\perp}f(tn+u)\,A_\varphi(u)\dd u\quad(t>0).
\]
Since $f$ is continuous off the origin, $K_\varphi$ is well defined and
continuous on $(0,\infty)$.

\begin{proposition}\label{prop:spectral}
There is a positive Borel measure $\rho_\varphi$ on $[0,\infty)$ with
$\int(1+E)^{-1}\rho_\varphi(\dd E)<\infty$ and
\begin{equation}\label{eq:laplace}
 K_\varphi(t)=\int_{[0,\infty)}e^{-tE}\,\rho_\varphi(\dd E),\qquad t>0 .
\end{equation}
\end{proposition}

\begin{proof}
Let $a_v=|v|$, which is the cell area of $\Gamma_v$
(Remark~\ref{rem:meshes}), put $c_R(u)=a_vR^{-2}\varphi(u/R)$ for
$u\in\Gamma_v\subset n^\perp$, $\Delta t=2|v|/R$, $t_m=m\Delta t$, and
\[
 K_R(m)=\frac{R^3}{\chi_R}\sum_{u,w\in\Gamma_v}\overline{c_R(u)}\,c_R(w)\,
 h(2mv+w-u)\qquad(m\ge d).
\]
The element $-\vt_v\in\Bthree$ fixes $v$ and reverses $w-u\in v^\perp$,
so the kernel $(u,w)\mapsto h(2mv+w-u)$ is symmetric. Hence the
real--imaginary cross terms cancel, and $K_R(m)$ is the sum of the two
quadratic forms in the real and imaginary parts of $c_R$. By
\textup{(M$_d$)}, $K_R(m)=\int_{[0,1]}\lambda^{m-d}\nu_R(\dd\lambda)$ for a
finite positive measure $\nu_R$. Let $\rho_R$ be the image of
$\nu_R|_{(0,1]}$ under $\lambda\mapsto-\log\lambda/\Delta t$; then
$K_R(m)=\int e^{-\tau_mE}\rho_R(\dd E)$ for $m>d$, with
$\tau_m=t_m-t_d$.

If $m=m(R)$ is chosen with $t_m\to t>0$, then $\tau_m\to t$, and the
points $(2mv+w-u)/R$ with $u,w\in R\operatorname{supp}\varphi$ converge,
uniformly, to $tn+(w'-u')$ with $u',w'\in\operatorname{supp}\varphi$; they
stay in a compact set avoiding the origin. By
Proposition~\ref{prop:spatial}(a) and two Riemann sums on
$R^{-1}\Gamma_v$, whose cells have area $a_v/R^2$,
\begin{equation}\label{eq:KRlimit}
 K_R(m)\longrightarrow\iint\overline{\varphi(u')}\varphi(w')f(tn+w'-u')\dd u'\dd w'
 =K_\varphi(t).
\end{equation}

Fix $\gamma>0$ and choose $m(R)$ with $\tau_{m(R)}\to\gamma/2$. Eventually
$\gamma-\tau_{m(R)}\ge\gamma/4$, so the measures $e^{-\gamma E}\rho_R$ have total
mass at most $K_R(m(R))$ and mass at most $e^{-\gamma M/4}K_R(m(R))$ on
$[M,\infty)$. They are therefore bounded and tight. Extract weak limits
$\rho^{(\gamma)}$ simultaneously for $\gamma=2^{-k}$, $k\in\N$. Since
$e^{-(\gamma'-\gamma)E}$ is bounded and continuous,
$\rho^{(\gamma')}=e^{-(\gamma'-\gamma)E}\rho^{(\gamma)}$ for $\gamma<\gamma'$, so
$\rho_\varphi=e^{\gamma E}\rho^{(\gamma)}$ is a well-defined locally finite
measure. Given $t>0$ take $\gamma<t$ in this list; the functions
$e^{-(\tau_m-\gamma)E}$ converge uniformly on $[0,\infty)$ to
$e^{-(t-\gamma)E}$, which together with \eqref{eq:KRlimit} gives
\eqref{eq:laplace}. The measure $\rho_\varphi$ may depend on the packet
and on the extraction; the left side of \eqref{eq:laplace} does not.

Finally $\int_0^1|K_\varphi(t)|\dd t\le\|A_\varphi\|_\infty\int_{|y|\le c}f<\infty$,
and integrating \eqref{eq:laplace} over $t\in(0,1)$ gives
$\int\frac{1-e^{-E}}E\,\rho_\varphi(\dd E)<\infty$, which is equivalent to
$\int(1+E)^{-1}\rho_\varphi(\dd E)<\infty$.
\end{proof}

\subsection{A Fourier density on punctured frequency space}\label{sec:fourier}
We use $\widehat\phi(x)=\int e^{-ix\cdot p}\phi(p)\dd p$.

\begin{proposition}\label{prop:fourier}
There is a real continuous function $C$ on $\R^3\setminus\{0\}$, bounded
on each set $\{|p|\ge\epsilon\}$, with $|C(p)|\le C'|p|^{-3}$ for
$|p|\le1$ and $C\circ P=C$ for $P\in\Bthree$, such that
\begin{equation}\label{eq:fourier}
 \int\widehat\phi\dd\mu=\int C(p)\,\phi(p)\dd p
 \qquad\text{for every Schwartz }\phi\text{ with }\phi(0)=0 .
\end{equation}
\end{proposition}

\begin{proof}
The atom contributes the constant $c_0$, and $f\one_{\{|x|\le1\}}$ is
integrable, so its Fourier transform is bounded and continuous. For the
remainder use a smooth partition of unity
$\{\zeta_{L,q}\}$ of $\{|x|\ge1\}$, where $L=2^i$, $i\ge0$, each
$\zeta_{L,q}(x)=\zeta_{1,q}(x/L)$ is supported in a shell
$L\le|x|\le4L$ and in one of finitely many closed cones $\mathcal C_q$,
and $q$ ranges over a fixed finite set. The cones are chosen as follows:
after a signed permutation the dominant coordinate is $x_1>0$, and the
three linearly independent vectors $b_1=e_1$, $b_2=e_1+s_2e_2$,
$b_3=e_1+s_3e_3\in\mathcal D$, with $s_i$ the sign of $x_i$, satisfy
$x\cdot b_i\ge|x|/\sqrt3$. These inequalities persist, with
$|x|/2$ in place of $|x|/\sqrt3$, on slightly larger cones
$\mathcal C'_q\supset\mathcal C_q$.

On $\mathcal C'_q\cap\{L/2\le|x|\le8L\}$, Proposition~\ref{prop:spatial}(c)
gives $|f|\le C$ and $|\Delta^j_{tb_i}f|\le C(t/L)^j$ for $j\le4$ and
$0<t\le c'L$. By the discrete product rule, the same bounds hold for
$f_{L,q}=\zeta_{L,q}f$ on all of $\R^3$, with $c'$ small enough that all
stencils starting in the support stay in the larger region; and the
support has volume $O(L^3)$. For $p\ne0$ choose $i$ with
$|p\cdot b_i|\ge c|p|$. If $L|p|\ge\pi/(cc')$ put
$t=\pi/|p\cdot b_i|\le c'L$. The Fourier multiplier of $\Delta^4_{tb_i}$
is $(e^{itp\cdot b_i}-1)^4$, of modulus $16$, so
\[
 |\widehat{f_{L,q}}(p)|\le\tfrac1{16}\|\Delta^4_{tb_i}f_{L,q}\|_1\le CL^{-1}|p|^{-4}.
\]
Otherwise $|\widehat{f_{L,q}}(p)|\le\|f_{L,q}\|_1\le CL^3$. Hence
$\sum_{L,q}|\widehat{f_{L,q}}(p)|\le C\sum_L\min(L^3,L^{-1}|p|^{-4})$,
which converges uniformly on $\{|p|\ge\epsilon\}$ and is $O(|p|^{-3})$
as $p\to0$. Define $C$ as $c_0$ plus the transform of the inner part plus
$\sum_{L,q}\widehat{f_{L,q}}$. Then $C$ is continuous off the origin
with the stated bounds.

For \eqref{eq:fourier}, $\int\widehat\phi f_{L,q}=\int\phi\,\widehat{f_{L,q}}$
for each piece. On the spatial side the pieces sum to $\int\widehat\phi\dd\mu$
by rapid decay of $\widehat\phi$ and the polynomial bound
$\mu([-L,L]^3)\le CL^{a}$, which follows from $\chi_{2R}\le C\chi_R$
(Lemma~\ref{lem:block}). On the frequency side, $|\phi(p)|\le C|p|$
near zero, so $|\phi|\sum|\widehat{f_{L,q}}|$ is dominated by an
integrable function. Since $\mu$ is even and $\Bthree$-invariant, $C$ is
real and $\Bthree$-invariant.
\end{proof}

\section{Angular rigidity}\label{sec:angular}

This section carries out Steps~2 and~3 of Section~\ref{sec:idea}. We first
combine the spectral representation and the Fourier density of
Section~\ref{sec:limits} into the statement that the Fourier transform of a
tangent is a Stieltjes function on every line parallel to one of the two
normals (Lemma~\ref{lem:stieltjes}). The complex-angle argument then shows
that the tangent is rotationally invariant (Theorem~\ref{thm:rotation}).

\subsection{Line restrictions are Stieltjes functions}
Let $v$, $n$ be as in Section~\ref{sec:spectral} and write frequencies
as $\omega n+k$ with $\omega\in\R$, $k\in n^\perp$. For
$\varphi\in C^\infty_c(n^\perp;\C)$ let
$\widehat\varphi(k)=\int_{n^\perp}e^{-ik\cdot w}\varphi(w)\dd w$.

\begin{lemma}\label{lem:line}
For every $\omega\ne0$,
\begin{equation}\label{eq:line}
 \int_{n^\perp}|\widehat\varphi(k)|^2\,C(\omega n+k)\,\frac{\dd k}{(2\pi)^2}
 =c_0\|\varphi\|_2^2+\int_{(0,\infty)}\frac{2E}{\omega^2+E^2}\,\rho_\varphi(\dd E).
\end{equation}
\end{lemma}

\begin{proof}
Let $\psi\in C^\infty_c(\R\setminus\{0\})$ be real and apply
\eqref{eq:fourier} to $\phi(\omega n+k)=\psi(\omega)|\widehat\varphi(-k)|^2/(2\pi)^2$,
which vanishes at the origin. Since
$A_\varphi(u)=\int e^{ik\cdot u}|\widehat\varphi(k)|^2\dd k/(2\pi)^2$,
we have $\widehat\phi(tn+u)=\widehat\psi(t)A_\varphi(u)$. The reflection
$\vt_v\in\Bthree$ fixes $n^\perp$ and reverses $n$, so
$\int f(tn+u)A_\varphi(u)\dd u=K_\varphi(|t|)$ for $t\ne0$, and the
spatial side of \eqref{eq:fourier} equals
$c_0\|\varphi\|_2^2\widehat\psi(0)+\int_\R\widehat\psi(t)K_\varphi(|t|)\dd t$;
Fubini's theorem applies because $K_\varphi$ is integrable near $0$
(Proposition~\ref{prop:spectral}) and $\widehat\psi$ decays rapidly.
Insert \eqref{eq:laplace}. For $E>0$ the Fourier transform of $e^{-E|t|}$
is $2E/(\omega^2+E^2)$; the exchange of the $t$- and $E$-integrals is
justified since $\int e^{-E|t|}|\widehat\psi(t)|\dd t$ is at most
$\|\widehat\psi\|_1$ for $E\le1$ and at most $2\|\widehat\psi\|_\infty/E$
for $E>1$, while $\int(1+E)^{-1}\dd\rho_\varphi<\infty$. The atom of
$\rho_\varphi$ at $E=0$ contributes $\int\widehat\psi=2\pi\psi(0)=0$. Thus
both sides of \eqref{eq:line}, with $|\widehat\varphi(-k)|^2$ on the
left, have the same integral against every such $\psi$. The element
$-\vt_v\in\Bthree$ maps $\omega n+k$ to $\omega n-k$, which allows us to
replace $|\widehat\varphi(-k)|^2$ by $|\widehat\varphi(k)|^2$. Both sides
are continuous in $\omega\ne0$, by Proposition~\ref{prop:fourier} and
dominated convergence, which gives equality pointwise.
\end{proof}

\begin{lemma}\label{lem:stieltjes}
Let $k_0\in n^\perp$. The function $S(Z)=C(\sqrt Z\,n+k_0)$, $Z>0$, has
the form
\begin{equation}\label{eq:stieltjes}
 S(Z)=b+\int_{[0,\infty)}\frac{\xi(\dd s)}{Z+s},\qquad b\ge0,\quad
 \int\frac{\xi(\dd s)}{1+s}<\infty .
\end{equation}
Consequently $S$ extends holomorphically to $\C\setminus(-\infty,0]$, and
there
\begin{equation}\label{eq:stieltjesbound}
 |S(Z)|\le S(1)\max\Bigl\{1,\ \sup_{s\ge0}\frac{1+s}{|Z+s|}\Bigr\};
 \qquad \sup_{s\ge0}\frac{1+s}{|Z+s|}\le 3+\frac4{|\Im Z|}\quad
 \text{if }|Z+1|\le\tfrac12 .
\end{equation}
\end{lemma}

\begin{proof}
Let $\varphi$ be a smooth compactly supported product packet with
$\|\varphi\|_2=1$ and put $\varphi_\ell(u)=\ell^{-1}e^{ik_0\cdot u}\varphi(u/\ell)$.
Then $|\widehat\varphi_\ell|^2\dd k/(2\pi)^2$ is a probability measure that
concentrates at $k_0$ as $\ell\to\infty$. For $\omega>0$ all frequencies
$\omega n+k$ have norm at least $\omega$, so by
Proposition~\ref{prop:fourier} the left side of \eqref{eq:line} tends to
$C(\omega n+k_0)$. Set $Z=\omega^2$ and $s=E^2$, and write the right
side as $\int_{[0,\infty]}\kappa_Z\dd\zeta_\ell$, where
$\kappa_Z(s)=(1+s)/(Z+s)$, $\kappa_Z(\infty)=1$, and $\zeta_\ell$ is the
image of $2E(1+E^2)^{-1}\one_{\{E>0\}}\rho_{\varphi_\ell}(\dd E)$ under
$E\mapsto E^2$ plus $c_0\delta_\infty$. Since $\kappa_1\equiv1$, the
total masses of $\zeta_\ell$ converge to $C(n+k_0)$ and are bounded. On the
compact space $[0,\infty]$ extract a weak limit $\zeta$; each $\kappa_Z$,
$Z>0$, is continuous there, so $S(Z)=\int\kappa_Z\dd\zeta$ for all
$Z>0$. Putting $b=\zeta(\{\infty\})$ and
$\xi=(1+s)\,\zeta|_{[0,\infty)}$ gives \eqref{eq:stieltjes}. The first
bound in \eqref{eq:stieltjesbound} follows from $|\kappa_Z|\le
\sup_s(1+s)/|Z+s|$ and $\zeta([0,\infty])=S(1)$. For the second,
$|Z+s|\ge|\Im Z|$ for $s\le3$, while $|Z+s|\ge s-\frac32\ge s/2$ for
$s>3$.
\end{proof}

The bound \eqref{eq:stieltjesbound} uses nothing but positivity of the
spectral measure on one line; it is uniform over all Stieltjes functions
with a given value at $Z=1$.

\subsection{The complex angle}
\begin{theorem}\label{thm:rotation}
Let $h$ be admissible and let $\mu$, $f$, $C$ be as above. Then
$C(Op)=C(p)$ for $O\in O(3)$ and $p\ne0$, $O_\#\mu=\mu$, and $f(Ox)=f(x)$
for $x\ne0$.
\end{theorem}

\begin{proof}
\emph{Step 1: the averaged angular profile.} Fix $z_0\ne0$, $a\in\R$ and
$\sigma>0$, let $\gamma_\sigma$ be the centred Gaussian density of
variance $\sigma^2$, and put
\[
 \Psi(\theta)=\int_\R\gamma_\sigma(v-a)\,
 C\bigl(e^v\cos\theta,e^v\sin\theta,z_0\bigr)\dd v,\qquad\theta\in\R .
\]
The integrand is bounded since $C$ is bounded on $\{|p|\ge|z_0|\}$. The
rotation by $\pi/2$ about $e_3$ belongs to $\Bthree$, so $\Psi$ has period
$\pi/2$.

\emph{Step 2: one chart.} Let $\alpha\in\{0,-\pi/4\}$ be the angle of a
normal $n$ in the $(e_1,e_2)$-plane ($n=e_1$, resp.\
$n=(e_1-e_2)/\sqrt2$), let $e$ be the horizontal unit vector with
$(n,e)$ positively oriented, and write $u=\theta-\alpha$. On an interval
$I_k=(k\pi/2,(k+1)\pi/2)$ of $u$ put $\epsilon=\sgn(\sin u)$ and
$s=\epsilon\sin u>0$, and change variables $w=v+\log s$. Since the
horizontal frequency is $e^v\cos u\,n+e^v\sin u\,e=e^w\cot u\,n+\epsilon e^w e$,
and $C$ is even in the normal component,
\begin{equation}\label{eq:chart}
 \Psi(\theta)=\int_\R\gamma_\sigma(w-a-\log s)\,S_{w,\epsilon}(\cot^2u)\dd w,
 \qquad S_{w,\epsilon}(Z)=C\bigl(e^w\sqrt Z\,n+\epsilon e^we+z_0e_3\bigr).
\end{equation}
For fixed $w$, $Z\mapsto S_{w,\epsilon}(e^{2w}Z)$ is the line restriction
of Lemma~\ref{lem:stieltjes} with transverse frequency
$k_0=\epsilon e^we+z_0e_3\in n^\perp$, and rescaling $Z$ preserves the
form \eqref{eq:stieltjes}; hence $S_{w,\epsilon}$ extends to
$\C\setminus(-\infty,0]$ with the bound \eqref{eq:stieltjesbound}, and
$S_{w,\epsilon}(1)\le\sup_{|p|\ge|z_0|}|C(p)|$ uniformly in $w$. The family
$\{S_{w,\epsilon}\}_w$ is locally bounded on the slit plane and depends
continuously on $w$ on the positive axis, so by Vitali's theorem it
depends continuously on $w$ on the whole slit plane.

Now let $u=x+iy$ range over the vertical strip $\Sigma_k=\{k\pi/2<x<(k+1)\pi/2\}$.
There $\Re s=\epsilon\sin x\cosh y>0$, so $\log s$ has a holomorphic
branch, real for $y=0$, with $|\Im\log s|<\pi/2$. Moreover
\begin{equation}\label{eq:cot}
 \cot(x+iy)=\frac{\sin2x-i\sinh2y}{\cosh2y-\cos2x},
\end{equation}
whose real part does not vanish in $\Sigma_k$; hence $\cot^2u\notin(-\infty,0]$
there. The integrand in \eqref{eq:chart} is therefore holomorphic in
$u\in\Sigma_k$, and on compact subsets of $\Sigma_k$ it is bounded by a
constant times $|\gamma_\sigma(w-a-\log s)|\le
(2\pi\sigma^2)^{-1/2}e^{\pi^2/(8\sigma^2)}e^{-(w-a-\Re\log s)^2/(2\sigma^2)}$.
Thus \eqref{eq:chart} defines a holomorphic function $\Psi_{\alpha,k}$ on
$\Sigma_k+\alpha$ which equals $\Psi$ on its real interval.

\emph{Step 3: gluing.} The strips $\Sigma_k+\alpha$, $k\in\Z$, omit the
lines $\Re\theta\in\alpha+\frac\pi2\Z$. For $\alpha=0$ and $\alpha=-\pi/4$
these two families of lines are disjoint, so every point of $\C$ lies in
a strip of one of the charts, and every nonempty intersection of a strip
of one chart with a strip of the other is a vertical strip of width
$\pi/4$, on which both functions equal $\Psi$ on the real segment. By the
identity theorem the functions $\Psi_{\alpha,k}$ glue to an entire function,
still denoted $\Psi$, (Figure~\ref{fig:strips}). Since $\Psi(\theta+\pi/2)-\Psi(\theta)$
vanishes on $\R$, $\Psi$ has period $\pi/2$.

\emph{Step 4: growth.} For each $x=\Re\theta$ one of the two values of
$\alpha$ satisfies $|\sin2(x-\alpha)|\ge1/\sqrt2$, since
$\max(|\sin2x|,|\cos2x|)\ge1/\sqrt2$. In that chart, by \eqref{eq:cot},
uniformly in such $x$, $\cot^2(u)\to-1$ as $|y|\to\infty$ and
\[
 |\Im\cot^2(x+iy)|=\frac{2|\sin2x\,\sinh2y|}{(\cosh2y-\cos2x)^2}\ge c\,e^{-2|y|}
 \qquad(|y|\ge1),
\]
with $x$ now denoting $\Re u$. By \eqref{eq:stieltjesbound}, for $|y|$
larger than some $Y_0$, $|S_{w,\epsilon}(\cot^2u)|\le Ce^{2|y|}$ uniformly
in $w$. The integral of $|\gamma_\sigma(w-a-\log s)|$ over $w$ equals
$\exp((\Im\log s)^2/(2\sigma^2))\le\exp(\pi^2/(8\sigma^2))$. For
$|y|\le Y_0$ we use continuity and periodicity. Altogether
\begin{equation}\label{eq:growth}
 |\Psi(\theta)|\le C_{z_0,\sigma}\,e^{2|\Im\theta|}\qquad(\theta\in\C).
\end{equation}

\emph{Step 5: the harmonics.} By periodicity,
$\Psi(\theta)=\sum_{m\in4\Z}a_me^{im\theta}$ with
\[
 a_m=\frac2\pi\int_0^{\pi/2}\Psi(\theta)\,e^{-im\theta}\dd\theta .
\] For $m\ne0$
shift the contour to $\Im\theta=-\sgn(m)\,Y$; the vertical sides cancel by
periodicity, and by \eqref{eq:growth} $|a_m|\le C_{z_0,\sigma}e^{(2-|m|)Y}$.
Since $|m|\ge4$, letting $Y\to\infty$ gives $a_m=0$. Thus $\Psi$ is constant.

\emph{Step 6: conclusion.} Letting $\sigma\downarrow0$, continuity and
boundedness of $C$ on $\{|p|\ge|z_0|\}$ show that
$\theta\mapsto C(e^a\cos\theta,e^a\sin\theta,z_0)$ is constant, for every
$a$ and every $z_0\ne0$; by continuity also for $z_0=0$. Hence $C$ is
invariant under rotations about $e_3$, and by $\Bthree$-invariance under
rotations about $e_1$ and $e_2$. These generate $SO(3)$, and $-I\in\Bthree$,
so $C\circ O=C$ for $O\in O(3)$.

Let $\phi$ be Schwartz and let $\beta$ be a radial Schwartz function with
$\beta(0)=1$. The functions $\phi-\phi(0)\beta$ and
$\phi\circ O-\phi(0)\beta$ vanish at the origin, their products with
$C$ are integrable (by $|C(p)|\le C'|p|^{-3}$), and their integrals
against $C$ agree by the change of variables $p\mapsto Op$. By
\eqref{eq:fourier}, $\int\widehat\phi\circ O\dd\mu=\int\widehat\phi\dd\mu$.
Since $\mu$ has polynomial growth and the Fourier transform is a
bijection of the Schwartz class, $O_\#\mu=\mu$. Uniqueness of the
continuous version of $f$ off the origin gives $f\circ O=f$.
\end{proof}

\subsection{\texorpdfstring{Lattice points; proof of Theorem~\ref{thm:main}(ii)}{Lattice points}}
\begin{corollary}\label{cor:points}
Let $h$ be admissible. Then for $O$, $x$, $R_j$, $k_j$, $\ell_j$ as in
Theorem~\ref{thm:main}(ii), $R_j^3[h(k_j)-h(\ell_j)]/\chi^h_{R_j}\to0$; the annular
version also holds. Moreover, for every tangent $\mu$ realized along
$R_j$ and every integer $s\ge1$, $\chi^h_{sR_j}/\chi^h_{R_j}\to\mu([-s,s]^3)$.
\end{corollary}

\begin{proof}
Every subsequence has a further subsequence along which
Proposition~\ref{prop:spatial} applies. By part (a) of that proposition
the normalized difference converges to $f(x)-f(Ox)$, which vanishes by
Theorem~\ref{thm:rotation}. In the annular version extract also
convergent subsequences $k_j/R_j\to x$, $\ell_j/R_j\to y$ with $|x|=|y|$.
The last assertion follows as for the unit cube, since
$\chi_{sR}/\chi_R=\mu_R([-s,s]^3)$.
\end{proof}

\begin{proof}[Proof of Theorem~\ref{thm:main}(ii)]
By Proposition~\ref{prop:admissible} a strongly reflection-positive kernel
is admissible; apply Lemma~\ref{lem:pointbound} and
Corollary~\ref{cor:points}.
\end{proof}

For a strongly reflection-positive kernel, Theorem~\ref{thm:main}(i) also
follows at once from Propositions~\ref{prop:admissible} and
\ref{prop:spatial} and Theorem~\ref{thm:rotation}, together with the
identity $\int(\psi\circ O-\psi)\dd\mu_R\to0$, which follows from
$O_\#\mu=\mu$ by the usual subsequence argument. The next section removes
\ref{K5}.

\section{\texorpdfstring{Site mirrors suffice: proof of Theorem~\ref{thm:main}(i)}{Site mirrors suffice}}\label{sec:site}

Without \ref{K5} the axial transfer operator $T_{e_1}$ is only a
contraction, and Lemma~\ref{lem:differences} controls only even steps
along the coordinate axes. A kernel satisfying \ref{K1}--\ref{K4} may
then oscillate between the two parity classes along an axis on the
microscopic scale, and \ref{O1} may fail. We remove this oscillation by
smoothing, without changing the tangents.

Let $\nu_1$ be the binomial weight $(1,4,6,4,1)/16$ on $\{-2,\dots,2\}$,
$\nu=\nu_1^{\otimes3}$ on $\{-2,\dots,2\}^3$, and
\[
 \bar g(z)=\sum_c\nu(c)\,g(z+c).
\]
Equivalently, $\nu=w*\tilde w$ with $w=\omega_1^{\otimes3}$,
$\omega_1=(\frac14,\frac12,\frac14)$ on $\{-1,0,1\}$, so
$\bar g(x-y)$ is the covariance of the averages
$\sum_a w(a)S_{x+a}$ whenever $g$ is the covariance of a field $S$.

\begin{lemma}\label{lem:smooth-basic}
Let $g$ satisfy \ref{K1}--\ref{K4}. Then $\bar g$ takes values in
$[0,1]$, is $\Bthree$-invariant and positive definite, and
$(3/8)^3g\le\bar g$. Moreover $\bar g$ satisfies \textup{(M$_1$)}.
\end{lemma}

\begin{proof}
The first assertions are immediate, using $\nu(0)=(6/16)^3$ and, for
positive definiteness,
$\sum c_ic_j\bar g(x_i-x_j)=\sum_{a,b}w(a)w(b)\sum c_ic_jg((x_i+a)-(x_j+b))$.
For (M$_1$) let $v\in\{e_1,e_1-e_2\}$, $\vt=\vt_v$, and $a$ be supported
on $\Gamma_v$. Since $w$ is invariant under $\vt$ and $\vt(nv)=-nv$,
\[
 \bar g\bigl(x-y+2(n+1)v\bigr)
 =\sum_{b,b'}w(b)w(b')\,g\bigl((x+v+b)-\vt(y+v+b')+2nv\bigr)
 \qquad(x,y\in\Gamma_v).
\]
Since $|b\cdot v|\le|v|^2$ for $b\in\{-1,0,1\}^3$, the points $x+v+b$
lie in $\Hp_v$. Hence
$\sum a(x)a(y)\bar g(x-y+2(n+1)v)=\ip A{T_v^{2n}A}_v$ with
$A=\sum_{x,b}a(x)w(b)[x+v+b]$, which is a moment sequence on $[0,1]$ by
Corollary~\ref{cor:moments}.
\end{proof}

The key point is a parity balance of the binomial weight.

\begin{lemma}\label{lem:parity}
Let $\beta_1(c)=\nu_1(c-1)-\nu_1(c)$ for $c\in\Z$, and
$\kappa_1=\frac1{16}(1,3,3,1)$ on $\{-2,-1,0,1\}$. Then
$\beta_1(c)=\kappa_1(c-2)-\kappa_1(c)$ for all $c$. Consequently, with
$\kappa=\kappa_1\otimes\nu_1\otimes\nu_1$, which satisfies
$\sum|\kappa|=\frac12$, every function $g$ satisfies
\[
 \Delta_{e_1}\bar g(x)=\sum_c\kappa(c)\,\Delta_{2e_1}g(x+c).
\]
\end{lemma}

\begin{proof}
$16\beta_1=(-1,-3,-2,2,3,1)$ on $c=-2,\dots,3$ and vanishes elsewhere,
and $16(\kappa_1(c-2)-\kappa_1(c))$ takes the same values. In particular
$\beta_1$ has zero sum on each parity class, which is the point. Since
$\Delta_{e_1}\bar g(x)=\sum_c\beta_1(c_1)\nu_1(c_2)\nu_1(c_3)g(x+c)$,
summation by parts in $c_1$ gives the identity.
\end{proof}

\begin{proposition}\label{prop:smooth-admissible}
If $g$ satisfies \ref{K1}--\ref{K4}, then $\bar g$ is admissible with
delay $1$.
\end{proposition}

\begin{proof}
(M$_1$) is Lemma~\ref{lem:smooth-basic}. By that lemma
$U_g\le(8/3)^3U_{\bar g}$. For \ref{O2}, note that
$\Delta^j_{2lv}\bar g=\sum_c\nu(c)\,\Delta^j_{2lv}g(\cdot+c)$, and $g$ satisfies \ref{O2} by
Proposition~\ref{prop:admissible}; the shifts $|c|\le2\sqrt3$ only change
constants. For \ref{O1} repeat the path argument in the proof of
Proposition~\ref{prop:admissible}. Steps along diagonal vectors are
controlled by \eqref{eq:diff} with $l=1$, since $\sigma(T_v)\subset[0,1]$
for diagonal $v$ (Lemma~\ref{lem:T}(ii)), and then averaged against
$\nu$. Steps along $\pm e_1$ are controlled by Lemma~\ref{lem:parity}
and \eqref{eq:diff} with the even step $2$, applied to $g$ at the points
$x+c$. In both cases the bound is $(C/r)U_g(r/C)\le(C'/r)U_{\bar g}(r/C)$.
\end{proof}

\begin{lemma}\label{lem:same-tangents}
Let $g$ satisfy \ref{K1}--\ref{K4}. Then $\chi^{\bar g}_R/\chi^g_R\to1$,
and for every $\psi\in C_c(\R^3)$
\[
 \int\psi\dd\mu^{\bar g}_R-\int\psi\dd\mu^g_R\longrightarrow0
 \qquad(R\to\infty).
\]
In particular $g$ and $\bar g$ have the same tangents along the same
sequences.
\end{lemma}

\begin{proof}
By Lemmas~\ref{lem:pointbound} (for $\bar g$) and \ref{lem:smooth-basic},
$U_g(r)\le C\chi^{\bar g}_r/r^3$. Since $\nu$ is supported in
$\|c\|_\infty\le2$, $|\chi^{\bar g}_R-\chi^g_R|$ is bounded by the sum of
$g$ over $\Lambda_{R+2}\setminus\Lambda_{R-2}$, which is at most
$CR^2U_g(R-2)\le C\chi^{\bar g}_R/R$. Next,
$\sum_x\bar g(x)\psi(x/R)=\sum_yg(y)\sum_c\nu(c)\psi((y-c)/R)$ by symmetry
of $\nu$, and $|\psi((y-c)/R)-\psi(y/R)|\le\omega_\psi(2\sqrt3/R)$ for the
modulus of continuity $\omega_\psi$. The mass of $\mu^g_R$ on the
$1$-neighbourhood of $\operatorname{supp}\psi$ is bounded
(Lemma~\ref{lem:block}), so the difference tends to zero.
\end{proof}

\begin{proof}[Proof of Theorem~\ref{thm:main}(i)]
Let $g$ be a reflection-positive kernel and $R_j\to\infty$. By
Proposition~\ref{prop:smooth-admissible} and
Proposition~\ref{prop:spatial} applied to $\bar g$, a subsequence of
$\mu^{\bar g}_{R_j}$ converges vaguely, and by
Lemma~\ref{lem:same-tangents} so does $\mu^g_{R_j}$, to the same limit.
Conversely every tangent of $g$ is a tangent of $\bar g$ along the same
sequence. By Theorem~\ref{thm:rotation} and
Proposition~\ref{prop:spatial}(b) for $\bar g$, every such tangent is
$O(3)$-invariant and has the form \eqref{eq:decomp}; moreover
$\mu([-1,1]^3)=1$ because $\mu^g_R([-1,1]^3)=1$ and the boundary is
$\mu$-null. Finally, if $\int(\psi\circ O-\psi)\dd\mu^g_R$ did not tend to
zero, a subsequence would stay away from zero and a further subsequence
would converge to a tangent $\mu$ with $\int(\psi\circ O-\psi)\dd\mu\ne0$,
contradicting invariance.
\end{proof}

\begin{remark}
Without \ref{K5} a lattice-point version of \eqref{eq:pointwise} still
holds on each of the eight classes of $\Z^3/2\Z^3$ separately, since
\ref{O1} holds for $g$ along even steps; the limiting profiles of the
eight classes coincide whenever an additional argument excludes a
macroscopic parity oscillation. One such argument, based on the infrared
bound, is given in Appendix~\ref{app:freecube}.
\end{remark}

\section{Verifying the hypotheses}\label{sec:models}

It remains to show that the two-point functions in
Theorems~\ref{thm:ising}, \ref{thm:potts} and~\ref{thm:sphere} are strongly
reflection-positive kernels. We first record two facts that hold for all
nearest-neighbour models: reflection positivity in finite boxes
(Lemma~\ref{lem:fvRP}), and the passage from finite boxes to kernels under
the common-limit condition (Proposition~\ref{prop:boxes}). The rest of the
section verifies the common-limit condition model by model.

\subsection{Two general facts}
We consider nearest-neighbour models on finite sets $D\subset\Z^3$ with
the induced edge set $E(D)$. Let $(\mathsf E,\lambda)$ be a compact
probability space of single-spin values, $K:\mathsf E\times\mathsf E\to(0,\infty)$
bounded, measurable and symmetric, and $Y:\mathsf E\to\R^m$ bounded and
measurable. The free Gibbs law on $D$ is
\[
 P_D(\dd\sigma)=Z_D^{-1}\prod_{\{x,y\}\in E(D)}K(\sigma_x,\sigma_y)
 \prod_{x\in D}\lambda(\dd\sigma_x),\qquad
 G_D(x,y)=E_D\bigl[Y(\sigma_x)\cdot Y(\sigma_y)\bigr].
\]
A uniform external field is included by tilting $\lambda$.

\begin{lemma}[Finite-volume reflection positivity]\label{lem:fvRP}
Let $\vt$ be a lattice translate of $\vt_v$, $v\in\mathcal D$, or of an
image of $\vtb$ under $\Bthree$, with closed positive half $\Hp$
(respectively $\{x_1\ge1\}$ for $\vtb$), and let $\vt D=D$.
\begin{enumerate}[label=\textup{(\alph*)},leftmargin=2em]
\item If $\vt$ is a site mirror, then
 $\sum_{i,j}c_ic_jG_D(x_i,\vt x_j)\ge0$ for all $x_i\in D\cap\Hp$ and
 $c_i\in\R$. No condition on $K$ is needed.
\item If $\vt$ is a bond mirror and
 $K(s,t)=\sum_\alpha\kappa_\alpha\phi_\alpha(s)\phi_\alpha(t)$ with
 $\kappa_\alpha\ge0$, bounded $\phi_\alpha$ and
 $\sum_\alpha\kappa_\alpha\|\phi_\alpha\|_\infty^2<\infty$, the same holds for
 $x_i\in D$ on the positive side.
\end{enumerate}
\end{lemma}

\begin{proof}
(a) A nearest-neighbour step changes $x\cdot v$ by at most one, for
axial and for diagonal $v$, so no edge joins the two open halves.
Given the spins on the mirror, the spins on the two open halves are
therefore conditionally independent, and $\vt$ interchanges their
conditional laws. With $F=\sum_ic_iY(\sigma_{x_i})$ and
$\vt F=\sum_ic_iY(\sigma_{\vt x_i})$,
$E_D[F\cdot\vt F]=E_D\bigl[|E_D[F\mid\sigma_{\rm mirror}]|^2\bigr]\ge0$.

(b) The weight factorizes as
$W(\sigma_+)\,W(\vt\sigma_-)\prod_yK(\sigma_y,\sigma_{\vt y})$, the
product running over the crossing edges $\{y,\vt y\}$. Expanding the
product,
\[
 Z_D\,E_D[F\cdot\vt F]=\sum_{\vec\alpha}\kappa_{\vec\alpha}
 \Bigl|\int F\,\Phi_{\vec\alpha}\,W\dd\lambda^{\otimes D_+}\Bigr|^2\ge0,
\]
with absolute convergence by the summability assumption.
\end{proof}

The kernels used below are
\begin{equation}\label{eq:crossing}
\begin{gathered}
 e^{\beta st}=\cosh\beta+\sinh\beta\,st,\qquad
 e^{\beta\one_{\{s=t\}}}=1+(e^\beta-1)\sum_{a=1}^q\one_a(s)\one_a(t),\\
 e^{\beta s\cdot t}=\sum_{\alpha\in\N_0^N}\frac{\beta^{|\alpha|}}{\alpha!}\,s^\alpha t^\alpha,
\end{gathered}
\end{equation}
for Ising, Potts and sphere spins, respectively; all have the form
required in (b) when $\beta\ge0$.

\begin{proposition}[From boxes to kernels]\label{prop:boxes}
In the setting above, assume $G_D\ge0$ and $|Y|\equiv1$, and that the
model is invariant under $\Bthree$. Suppose there is a function $g$ on
$\Z^3$ such that
\begin{enumerate}[label=\textup{(S)},leftmargin=2.5em]
\item[] \emph{Common-limit condition.}
\item for each of the mirrors $\vt_{e_1}$, $\vt_{e_1-e_2}$ and $\vtb$
 there are finite sets $D_n\uparrow\Z^3$ with $\vt D_n=D_n$ and
 $G_{D_n}(x,y)\to g(x-y)$ for all $x,y$, and for one mirror the $D_n$ are
 centred cubes.
\end{enumerate}
Then $g$ is a strongly reflection-positive kernel. If (S) is only known
for the two site mirrors, $g$ is a reflection-positive kernel.
Moreover, pointwise limits of (strongly) reflection-positive kernels are
(strongly) reflection-positive kernels.
\end{proposition}

\begin{proof}
\ref{K1}: $g\ge0$ and $g(0)=1$. \ref{K2}: the centred cubes are
$\Bthree$-invariant. \ref{K3}: each matrix $(G_D(x_i,x_j))$ is a Gram
matrix. \ref{K4}, \ref{K5}: Lemma~\ref{lem:fvRP} and passage to the
limit in finitely many terms. The last statement is clear, since every
condition involves finitely many values.
\end{proof}

The whole difficulty in the applications is thus hypothesis (S): one
needs a single limit for boxes adapted to different mirrors.

\subsection{\texorpdfstring{Potts and Ising: proof of Theorems~\ref{thm:ising} and~\ref{thm:potts}}{Potts and Ising}}
Let $q\ge2$, $\mathsf E=\{1,\dots,q\}$ with uniform $\lambda$,
$K(s,t)=e^{\beta\one_{\{s=t\}}}$, and let $Y(a)\in\R^q$ be the vertices of
the centred regular simplex normalized by $Y(a)\cdot Y(b)=(q\one_{\{a=b\}}-1)/(q-1)$.
Under the Edwards--Sokal coupling \cite{FK,ES,Grimmett} of $P_D$ with the
free random-cluster measure $\phi^{\rm f}_{D,p,q}$, $p=1-e^{-\beta}$, the
colours are constant on open clusters and independent and uniform on
distinct clusters. Since $\sum_aY(a)=0$,
\[
 G_D(x,y)=\phi^{\rm f}_{D,p,q}(x\leftrightarrow y)\ \ge0 .
\]
For $q=2$, $Y(\pm)=\pm1$, $Y\cdot Y=\sigma\sigma'$, and
$e^{\beta\one_{\{\sigma=\sigma'\}}}=e^{\beta/2}e^{(\beta/2)\sigma\sigma'}$;
so the $q=2$ kernel is the free Ising two-point function at $\beta/2$.
Theorem~\ref{thm:ising} is therefore the case $q=2$ of
Theorem~\ref{thm:potts}, combined with Theorem~\ref{thm:main}.

\begin{lemma}\label{lem:fk-mono}
If $D\subset D'$ are finite, the restriction to $E(D)$ of
$\phi^{\rm f}_{D',p,q}$ stochastically dominates $\phi^{\rm f}_{D,p,q}$
for $q\ge1$. In particular $D\mapsto\phi^{\rm f}_{D,p,q}(x\leftrightarrow y)$
is nondecreasing.
\end{lemma}

\begin{proof}
Under $\phi^{\rm f}_{D',p,q}$, conditionally on all other edges, an edge
$e$ is open with probability $p$ if its endpoints are joined by an open
path avoiding $e$, and with probability $p/(p+q(1-p))\le p$ otherwise.
This conditional probability is nondecreasing in the other edges, and,
for edges of $E(D)$, at least the corresponding conditional probability
under $\phi^{\rm f}_{D,p,q}$ given the same configuration on $E(D)$,
because open paths through $E(D')\setminus E(D)$ can only add
connections. Holley's criterion \cite{Holley} gives the domination; see
\cite[Ch.~4]{Grimmett}.
\end{proof}

\begin{proof}[Proof of Theorem~\ref{thm:potts}]
By Lemma~\ref{lem:fk-mono}, $G_D(x,y)$ is nondecreasing in $D$ and
bounded by one, so along every exhaustion it converges to
$\sup_DG_D(x,y)$, the supremum over finite $D\ni x,y$. This limit is
independent of the exhaustion, and by translation and $\Bthree$
covariance of finite volumes it has the form $g_{q,\beta}(x-y)$ with
$g_{q,\beta}$ invariant under $\Bthree$. Hence (S) holds for every
choice of mirror-symmetric boxes, and Proposition~\ref{prop:boxes} with
\eqref{eq:crossing} shows that $g_{q,\beta}$ is a strongly
reflection-positive kernel.

To identify $g_{q,\beta}(x-y)$ with $\phi^{\rm f}_{p,q}(x\leftrightarrow y)$:
by Lemma~\ref{lem:fk-mono} the laws $\phi^{\rm f}_{D,p,q}$ (with edges
outside $E(D)$ closed) increase with $D$, so their probabilities of
cylinder events of the form ``a given finite set of edges is open''
converge, and by inclusion--exclusion they converge weakly to a single
law $\phi^{\rm f}_{p,q}$. Let $C_M$ be the event that $x,y$ are joined
inside $\Lambda_M$ and $c_D=\phi^{\rm f}_{D,p,q}(x\leftrightarrow y)$.
For $D\supset\Lambda_M$, $c_{\Lambda_M}\le\phi^{\rm f}_{D,p,q}(C_M)\le c_D$.
Letting first $D\uparrow\Z^3$ and then $M\to\infty$, and using that
$C_M\uparrow\{x\leftrightarrow y\}$, gives
$\phi^{\rm f}_{p,q}(x\leftrightarrow y)=\lim_Dc_D$.
\end{proof}

\begin{remark}[Integrality of $q$]\label{rem:q1}
The proof uses the colour observable, not reflection positivity of
connectivity matrices, and the distinction matters. For bond percolation
at $p=\frac12$ on the four edges of a unit square, reflected in a
diagonal, take as marks the two vertices on the diagonal and one vertex
off it. Adjacent vertices are connected with probability $\frac9{16}$
and opposite ones with probability $\frac7{16}$, so the reflected
connectivity matrix is
$\frac1{16}\bigl(\begin{smallmatrix}16&7&9\\7&16&9\\9&9&7\end{smallmatrix}\bigr)$,
of determinant $-9/4096$. For this reason the argument is restricted to
integer $q\ge2$.
\end{remark}

\begin{remark}[The critical Ising state]
For the nearest-neighbour Ising model on $\Z^3$ the spontaneous
magnetization vanishes at $\beta_c$ \cite{ADS}, so the Gibbs state at
$\beta_c$ is unique \cite{LML} and coincides with the free state.
\end{remark}

For the Ising model the Messager--Miracle-Sol\'e inequalities \cite{MMS}
give in addition a relative comparison on scales where the cube mass grows.

\begin{proposition}[Ising: relative comparison]\label{prop:ising-relative}
Let $g$ be the free Ising two-point function at some $\beta\ge0$, and let
$\epsilon>0$. Along any sequence $R_j\to\infty$ with
$\chi_{2R_j}\ge(1+\epsilon)\chi_{R_j}$, and with $k_j,\ell_j$ as in
Theorem~\ref{thm:main}(ii), $g(k_j)/g(\ell_j)\to1$. At $\beta=\beta_c$, writing
$d_i=\chi_{2^{i+1}}/\chi_{2^i}$, for every $1<c<2$
\[
 \liminf_{J\to\infty}\frac{\#\{0\le i<J:\ d_i\ge c\}}J\ \ge\
 \frac{\log2-\log c}{\log8-\log c}>0 .
\]
\end{proposition}

\begin{proof}
Let $u_k=g(ke_1)$. By \cite{MMS} (see \cite[\S4.1]{ADC}),
$u_{\|v\|_1}\le g(v)\le u_{\|v\|_\infty}$ and $k\mapsto g(k,v_2,v_3)$ is
nonincreasing for $k\ge0$. The shell $\Lambda_{2R}\setminus\Lambda_R$ has
$O(R^3)$ sites and mass at least $\epsilon\chi_R$, and each term is at
most $u_R$; hence $u_R\ge c\epsilon\chi_R/R^3$. With $h=2\lfloor R/4\rfloor$,
Theorem~\ref{thm:main}(ii) gives $u_h\le C\chi_R/R^3$, and $u_{2h}\ge u_R$. By
Lemma~\ref{lem:T}(iii) and Corollary~\ref{cor:moments}, $(u_k)$ is a
moment sequence and so log-convex; hence
$u_{mh}\ge u_{2h}(u_{2h}/u_h)^{m-2}\ge c(\epsilon,m)\chi_R/R^3$ for each
fixed $m\ge2$. Choosing $m$ with $mh\ge\max(\|k_j\|_1,\|\ell_j\|_1)$
gives $R_j^3g(k_j)/\chi_{R_j}\ge c'>0$ and likewise for $\ell_j$, and
\eqref{eq:pointwise} gives the ratio limit.

For the second statement, coordinate monotonicity gives $d_i\le8$: round
each absolute coordinate up after division by two, keeping its sign;
each image has at most eight preimages. At $\beta_c$,
$g(x)\ge c(1+\|x\|_\infty)^{-2}$ (\cite[\S4]{ADC}, which rests on
\cite{Lieb}), so $\chi_R\ge c_1R$ and
$J^{-1}\sum_{i<J}\log d_i\ge\log2-O(J^{-1})$. Bounding each $\log d_i$ by
$\log c$ or $\log8$ according as $d_i<c$ or $d_i\ge c$ gives the claim.
\end{proof}

\subsection{\texorpdfstring{The XY model: Theorem~\ref{thm:sphere}(i)}{The XY model}}
Here $N=2$, $S_x=(\cos u_x,\sin u_x)$ with $u_x$ uniform on the circle,
and $Y=S$. For an edge $e=\{x,y\}$ write $c_e(u)=\cos(u_x-u_y)$ and
$s_e(u)=\sin(u_x-u_y)$.

\begin{lemma}[Ginibre]\label{lem:ginibre}
On a finite vertex set take edge couplings $0\le J_e\le K_e$. Then
$\langle\cos(\sum_xk_xu_x)\rangle_K\ge\langle\cos(\sum_xk_xu_x)\rangle_J\ge0$
for every finitely supported integer charge $k$; in particular
$\langle c_a\rangle_K\ge\langle c_a\rangle_J\ge0$.
\end{lemma}

\begin{proof}
Multiply the difference by $Z_KZ_J$ and write it as an integral over two
replicas $(u,v)$; substitute $(u-v,u+v)$, a surjective endomorphism of
the product torus preserving Haar measure. For $c_a$ the integrand becomes
\[
 2D_a(u)s_a(v)\exp\sum_e\bigl[(K_e+J_e)c_e(u)c_e(v)+(K_e-J_e)s_e(u)s_e(v)\bigr],
\]
and expanding the exponential gives a sum of squares with nonnegative
coefficients \cite{Ginibre}. The same substitution applies to
$\cos(\sum k_xu_x)$; nonnegativity is the case $J=0$.
\end{proof}

\begin{proof}[Proof of Theorem~\ref{thm:sphere}\textup{(i)}]
Adding a vertex as an isolated site does not change the law of the
others, and then turning on its edges increases every real Fourier
coefficient by Lemma~\ref{lem:ginibre}. Hence these coefficients are
nondecreasing along any exhaustion by induced subgraphs, and converge to
their supremum over finite sets. The imaginary coefficients vanish by
reflection of all angles, and the characters determine the finite-dimensional
laws; so the free laws converge weakly to one translation- and
$\Bthree$-invariant law, which is Gibbs because the finite-window
conditional kernels converge uniformly in the boundary spins. In
particular (S) holds, and Proposition~\ref{prop:boxes} with
\eqref{eq:crossing} shows that $g(x)=\E[S_0\cdot S_x]$ is strongly
reflection positive. Since a fixed inner volume at
$\beta_n\le\beta$ eventually bounds the coefficients from below and the
limit at $\beta$ bounds them from above, the same law is obtained when
$\beta_n\uparrow\beta$ at an arbitrary rate relative to the volume.
\end{proof}

\subsection{\texorpdfstring{The Heisenberg model: Theorem~\ref{thm:sphere}(ii)}{The Heisenberg model}}\label{sec:heis}
Let $N=3$, $e\in\Sph^2$, and let $P^e_{D,h}$ denote the free law on $D$ in
the uniform field $he$, i.e.\ with $\lambda$ replaced by
$\lambda_h(\dd s)\propto e^{h\,e\cdot s}\lambda(\dd s)$. By rotation
invariance we take $e=e_1$.

\begin{lemma}\label{lem:heis-positive}
Let $\beta\ge0$ and $h>0$. Along centred cubes, along their translates,
and along the bond-centred rectangles $[-L,L+1]\times[-L,L]^2$, the free
laws $P^e_{D,h}$ converge weakly to one law $P^e_{\beta,h}$, which is
translation- and $\Bthree$-invariant and Gibbs in the field $he$. Its dot
correlation $g_h$ is a strongly reflection-positive kernel.
\end{lemma}

\begin{proof}
\emph{Large field.} At one site, with the neighbours fixed, the
conditional law has density proportional to $e^{w\cdot s}$ with
$w=he+\beta\sum_{y\sim x}S_y$, $|w|\ge h-6\beta$. The covariance of this
law has norm at most $1/|w|$, and changing one neighbour moves $w$ by at
most $2\beta$; hence the total-variation influence of one neighbour is at
most $\beta/\sqrt{h-6\beta}$. For $h>6\beta+36\beta^2$ the six
influences sum to less than one, Dobrushin's condition holds
\cite{Dobrushin,Georgii}, and all three families of free volumes converge
to the unique Gibbs state in the field $he$.

\emph{Analytic continuation.} By the finite-volume Lee--Yang theorem for
isotropic ferromagnetic $\Sph^2$-spins (\cite[Prop.~6]{FRLY}, see also
\cite{DunlopNewman}), $Z_D(H)\ne0$ for site-dependent complex fields
with $\Re H_x^1>|H_x^2|+|H_x^3|$ at every site. Fix a compact
$K\subset\{\Re z>0\}$ and a finite list of marked site coordinates.
Using the elementary bound $\Re f'(0)/f(0)\ge-2c$ for entire functions
$f$ of exponential type $c$ without zeros in $\{\Re z\ge0\}$, applied to
the partition function as a function of one additional site field in
three slightly tilted frames, one obtains $\Re\langle u\cdot S_x\rangle\ge-2$
on a complex neighbourhood of the segment from a large real field to
$K$; Schwarz's lemma then bounds the one-point functions uniformly in
$D$, and Cauchy's estimates applied to iterated one-site derivatives
bound every fixed coordinate moment on $K$ uniformly in $D$. Thus the
finite-volume moments form normal families on $\{\Re z>0\}$. Any two
subsequential limits along the three families agree on the large real
interval, hence on the half-plane by the identity theorem. Compactness
of the spin space and density of polynomials identify one limit law at
every $h>0$; translating the exhaustion proves stationarity.

\emph{Kernel.} Expanding the bond and field factors, every integrated
monomial in the numerator of $\langle S^i_xS^i_y\rangle_{D,h}$ is a sphere
integral of a monomial with nonnegative coefficient, which vanishes if
some exponent is odd and is nonnegative otherwise. Hence $G_D\ge0$. The
field contributes identical single-site weights on both sides of every
mirror. Proposition~\ref{prop:boxes} applies.
\end{proof}

\begin{proposition}\label{prop:heis-ident}
Let $\beta\ge0$. As $h\downarrow0$, $P^e_{\beta,h}$ converges weakly to a
zero-field Gibbs state $Q^{e}_\beta$. The same law is the weak limit,
along any exhaustion by finite induced sets $D_n$, of the laws with all
exterior spins fixed to $e$ (the crossing bonds retained) and bulk field
$h_ne$, for any $h_n\ge0$ with $h_n\to0$.
\end{proposition}

\begin{proof}
Write $v_i=e\cdot S_i$, $L_{ij}=v_iv_j$, $G_{ij}=S_i\cdot S_j$ and
$T_{ij}=(t\cdot S_i)(t\cdot S_j)$ for a fixed unit $t\perp e$. Tilt a finite
window by $\exp(\sum a_iv_i+\sum_{i<j}b_{ij}G_{ij})$ with
$a_i,b_{ij}\ge0$. For ferromagnetic isotropic couplings and nonnegative
longitudinal fields, the means of $v_i$, $L_{ij}$ and $-T_{ij}$ are
nondecreasing in every coupling and every longitudinal field
\cite[Prop.~8]{KPV}; the tilt preserves these hypotheses. (For the
single-spin assertion condition on the amplitudes $|v_i|$: the signs form
an Ising ferromagnet, and the amplitudes are positively associated, as in
\cite{KPV}.) Enlarging the domain with fixed exterior $e$ decreases each
of the three tilted means: send the longitudinal fields of the added
spins to infinity, so that their priors converge to $\delta_e$ and each
bond to a pinned site supplies its original boundary term. Hence, for
any of the tilted means $U(D,h)$,
\[
 \lim_{n}U(D_n,h_n)=\inf_D\inf_{t>0}U(D,t)=\lim_{t\downarrow0}\inf_DU(D,t),
\]
by comparison with each fixed $(D,t)$ and continuity of each
finite-volume mean at $t=0$.

At fixed $h>0$ the pinned and the free limits agree: the crossing bonds
produce site fields $h+c_i$ with $0\le c_i\le6\beta$, and the argument of
Lemma~\ref{lem:heis-positive} applies to these offsets, with the large-field
interval again identifying the limits. Therefore every candidate limit
state, of either construction, has the same tilted means of $v_i$ and of
$G_{ij}=L_{ij}+2T_{ij}$ (axial symmetry about $e$). These are the partial
derivatives of the logarithmic Laplace transform of $(v_i,G_{ij})_{i,j}$,
so integration along rays identifies that transform for all
$a,b\ge0$, and hence the joint law of these bounded variables. They form
the Gram matrix of the labelled list $(e,(S_i)_i)$; equal Gram matrices
give an isometry of the two spans fixing $e$, which extends to an
element of the stabilizer of $e$ in $O(3)$. Every continuous cylinder
test averaged over this stabilizer factors through the Gram matrix, and
all candidate laws are stabilizer-invariant, so they agree on cylinder
tests. Compactness gives the asserted full limits.
\end{proof}

\begin{proof}[Proof of Theorem~\ref{thm:sphere}\textup{(ii)}]
By Lemma~\ref{lem:heis-positive}, each $g_h$ is a strongly
reflection-positive kernel; $g_{Q^e_\beta}=\lim_{h\downarrow0}g_h$ by
Proposition~\ref{prop:heis-ident}, so it is one as well
(Proposition~\ref{prop:boxes}). $Q^e_\beta$ is a zero-field Gibbs state,
since the finite-window kernels converge uniformly in the boundary spins
as $h\to0$.
\end{proof}

\begin{remark}
Without Proposition~\ref{prop:heis-ident}, the same proof shows that every
weak accumulation point of $P^e_{\beta,h_j}$, $h_j\downarrow0$, is a
zero-field Gibbs state whose two-point function is a strongly
reflection-positive kernel.
\end{remark}

\subsection{\texorpdfstring{All $N$: Theorem~\ref{thm:sphere}(iii), (iv)}{All N}}
\begin{proof}[Proof of Theorem~\ref{thm:sphere}\textup{(iv)}]
Weak limits of free-boundary laws along any exhaustion are Gibbs, since
the finite-window conditional equations hold once the domain contains
the window and its neighbours. If the Gibbs state is unique, every such
limit is that state, so (S) holds for every family of mirror-symmetric
boxes, and $G_D\ge0$ by the monomial expansion. Proposition~\ref{prop:boxes}
applies. For uniqueness when $6\tanh\beta<1$: if two neighbour
configurations differ at one site, the two single-site conditional
densities $w,v$ satisfy $w(x)v(y)\le e^{4\beta}v(x)w(y)$, which gives
total-variation distance at most $\tanh\beta$, and Dobrushin's criterion
\cite{Dobrushin} applies with six neighbours.
\end{proof}

The same argument applies to any weak limit of the unique states along
$\beta_j\uparrow\beta_{\rm u}$, where $\beta_{\rm u}\in(0,\infty)$ is the
first inverse temperature with more than one Gibbs state: each kernel is
strongly reflection positive, and Proposition~\ref{prop:boxes} passes
this to the limit, which is Gibbs at $\beta_{\rm u}$. That
$\beta_{\rm u}<\infty$ for $N\ge2$ follows from \cite{FSS}.

Theorem~\ref{thm:sphere}(iii) is proved in Appendix~\ref{app:selected}. Its
proof selects a state by a variational principle after adding
positive-definite auxiliary interactions that preserve reflection
positivity on suitably adapted tori, and then removes them. We record one
more consequence of Theorem~\ref{thm:main}, which needs no selection beyond
stationarity.

\begin{corollary}\label{cor:freecube}
Let $N\ge1$, $\beta\ge0$, and suppose that along one sequence of centred
free cubes $\Lambda_{L_m}$ the two-point functions converge,
$\langle S_x\cdot S_y\rangle^{\rm f}_{\Lambda_{L_m}}\to g(x-y)$, to a
function of $x-y$ only. Then $g$ is a reflection-positive kernel, and
Theorem~\ref{thm:main} applies.
\end{corollary}

\begin{proof}
Centred cubes are symmetric about both site mirrors $x_1=0$ and
$x_1=x_2$; apply Proposition~\ref{prop:boxes} with the two site mirrors.
\end{proof}

In this situation the lattice-point comparison \eqref{eq:pointwise} also
holds; see Appendix~\ref{app:freecube}.

\section{Open problems}\label{sec:remarks}

\begin{enumerate}[label=\textup{(\arabic*)},leftmargin=2em]
\item \emph{Scale invariance.} Is the tangent of the critical
 two-point function unique, and is its density homogeneous,
 $f(x)=c|x|^{-(1+\eta)}$? By Corollary~\ref{cor:points} this amounts to
 the convergence of $\chi_{sR}/\chi_R$ for every $s>0$, to a power of $s$.
 Rotational invariance and scale invariance together would determine the
 critical two-point function up to its exponent.
\item \emph{Higher correlations.} The method applies to any scalar
 kernel with \ref{K1}--\ref{K4}, for instance to the two-point function
 of a fixed local observable. For a Gaussian tangent field, Wick's
 formula transfers rotational invariance to all moments. Are the
 $n$-point functions, $n\ge3$, of the critical three-dimensional Ising
 model rotationally invariant in the scaling limit, and are they
 conformally covariant?
\item \emph{States.} Does the state $Q^e_\beta$ of
 Theorem~\ref{thm:sphere}(ii) coincide with the free state at $\beta_c$, that
 is, is $m_*(\beta_c)=0$ for the three-dimensional Heisenberg model?
 Which boundary condition produces the state of
 Theorem~\ref{thm:sphere}(iii)? Any state satisfying the common-limit
 condition (S) for the three mirrors falls under Theorem~\ref{thm:main}(ii).
\item \emph{Other models.} The random-cluster model with nonintegral $q$
 (Remark~\ref{rem:q1}) and the self-avoiding walk, the formal limit
 $N\to0$ of the $O(N)$ model, require a different positivity structure:
 for the latter, the limits $N\to0$, $|\Lambda|\to\infty$ and fugacity
 $\to z_c$ would have to be interchanged.
\item \emph{Unbounded spins.} Theorem~\ref{thm:main} applies to the
 normalized two-point function $g/g(0)$ of any reflection-positive
 translation-invariant state, bounded or not. For lattice $\varphi^4$
 models and other models of the Griffiths--Simon class, Griffiths'
 inequalities and reflection positivity are available, but the
 verification of the common-limit condition in Section~\ref{sec:models}
 is written for compact spin spaces. Does Theorem~\ref{thm:ising} hold for
 the critical lattice $\varphi^4_3$ model?
\item \emph{Dimension.} The angular argument works in coordinate planes,
 with the remaining frequencies as spectators, and the Fourier estimates
 of Proposition~\ref{prop:fourier} have analogues in every dimension.
 Does the proof extend to $\Z^d$? For $d=2$ this would give rotational
 invariance of planar two-point tangents without integrability.
\item \emph{Mechanism.} Which pairs of families of lattice mirrors
 produce charts whose boundaries interlace, and what growth exponent do
 the resulting entire functions have? On the cubic lattice the two site
 families give exponent $2$ against the harmonic gap $4$.
\end{enumerate}

\appendix
\section{\texorpdfstring{The selected state for general $N$}{The selected state for general N}}\label{app:selected}

We prove Theorem~\ref{thm:sphere}(iii). Fix $N\ge1$ and $\beta\ge0$. For a
stationary state with two-point function $g$, let $\sigma_g$ be the spectral
measure of $g$ on $\mathbb T^3$; it is positive of mass $g(0)=1$.

\emph{Auxiliary covariances.} Let $A$ be the symmetric walk matrix on
$\Z^3$ with weights $a,b,c>0$ on nearest-neighbour, face-diagonal and
body-diagonal steps, with $a>4b+4c$, $b>2c$, and let $M>6a+12b+8c$. The
covariance $C=(M-A)^{-1}$ is entrywise nonnegative and summable by its
Neumann series, and with $(u,v,w)=(\cos k_1,\cos k_2,\cos k_3)$ and
elementary symmetric polynomials $e_1,e_2,e_3$ of $u,v,w$,
$\widehat C=(M-2ae_1-4be_2-8ce_3)^{-1}$. Its periodization is the
covariance of the corresponding Gaussian field on a torus.

For the coordinate mirrors use even cubic tori. A coordinate site mirror
has no step between its two strict halves; at a coordinate bond
interface the reflected cross matrix of the Gaussian log weight has
transverse multiplier $a+2b(\cos p+\cos q)+4c\cos p\cos q>0$. For the
normal $(1,1,0)$ use the torus with periods $L(1,1,0)$, $L(1,-1,0)$,
$L(0,0,1)$, $L>4$ even; the layers $x_1+x_2\in\{0,L\}\bmod2L$ are fixed by
the reflection, and between the strict halves the only crossing steps are,
up to reversal, $(1,1,0)$ and $(1,1,\pm1)$, with interface matrix
$bI+c(T_3+T_3^*)>0$, $T_3$ the shift in direction $3$. In each case,
conditionally on the fixed spins, a positive semidefinite reflected cross
coupling is a sum of products of a linear form and its reflection with
nonnegative coefficients; expanding its exponential gives a sum of squares.
Consequently, adding any positive multiple of $C_{\rm per}$ to the
nearest-neighbour coupling of the sphere spins (component by component in
the dot product) preserves reflection positivity on the indicated torus.

\emph{Admissible variations.} Take a dense countable family $C_j$ from the
open parameter domain and $\alpha_j>0$ with
$\sum_j\alpha_j(1+\|C_j\|_1)<\infty$, and set
$J_\delta=J_{\rm nn}+\delta\sum_j\alpha_jC_j$, $\delta>0$, where
$J_{\rm nn}$ has weight $\beta$ on nearest-neighbour steps. Fix a weak
limit $\mu$ of the periodic Gibbs laws for $J_\delta$ on cubic tori of
side $2(n+1)$. It is invariant under translations, $\Bthree$ and internal
$O(N)$, and satisfies the two axial mirror inequalities (site and bond)
by Lemma~\ref{lem:fvRP}; $0\le g_\mu\le1$ by the monomial expansion.

It remains to obtain the diagonal-mirror inequality for the \emph{same}
$\mu$. Fix the normal $(1,1,0)$ and let $G$ be the group generated by the
coordinate sign changes and the swap of coordinates $1,2$. Varying the
mass, the weight on the steps $\pm e_3$, or the weight on the four
face-diagonal steps in the $(1,2)$-plane of one summand preserves the
strict inequalities above. If $C_j^{-1}-tD_j$ is the perturbed inverse
covariance, the resolvent identity gives
$C_j(t)-C_j-tC_jD_jC_j=t^2C_jD_jC_jD_jC_j(t)$, so
$J(t)=J_\delta+tH+O_{\ell^1}(t^2)$ for every $H$ whose Fourier transform is
a finite real combination of $\widehat C_j^2$, $w\widehat C_j^2$,
$uv\widehat C_j^2$. These span a uniformly dense subspace of the
continuous real $G$-invariant functions on $\mathbb T^3$: a signed measure
annihilating all $\widehat C_j^2$ annihilates the whole parameter family,
expansion at large $M$ gives all moments of $2ae_1+4be_2+8ce_3$, varying
$(a,b,c)$ gives all monomials in the $e_i$, which separate cubic orbits;
and every $G$-invariant cosine polynomial is a combination of $1,w,uv$
with cubic-invariant coefficients, using $u+v=e_1-w$,
$uv=e_2-e_1w+w^2$ and $w^3=e_1w^2-e_2w+e_3$.

\emph{Transfer of the mirror tests.} For a finite real packet $(d_x)$ on
the positive half of the diagonal mirror $\vt$, the desired
inequality is $\int T\dd\sigma_{g_\mu}\ge0$ with
$T(k)=\sum_{x,y}d_xd_y\cos(k\cdot(x-\vt y))$. The $G$-average $\bar T$
has the same integral in every $G$-invariant state. Given $\epsilon>0$
choose $H$ as above with $\|\widehat H+\bar T\|_\infty<\epsilon$, and let
$\mu_t$ be a thermodynamic limit for $J(t)$ on the diagonal-adapted tori;
it is stationary, $G$-invariant and satisfies the diagonal inequality. Let
$\mathcal E_J(\tau)=\frac12\sum_zJ(z)g_\tau(z)$ be the pair term of the
pressure variational principle \cite[\S15.4]{Georgii}. Both sequences of
periodic fundamental domains are van Hove, and finite-range approximation
changes each pressure per site by at most half the $\ell^1$ tail, so the
periodic limits maximize the same functional at their interactions.
Comparing $\mu_t$ and $\mu$ at $J(t)$ and at $J_\delta$ and adding, the
entropies cancel:
$\mathcal E_{(J(t)-J_\delta)/t}(\mu_t)\ge\mathcal E_{(J(t)-J_\delta)/t}(\mu)$
for $t>0$. A weak limit $\mu_H$ as $t\downarrow0$ gives
$\int\widehat H\dd\sigma_{g_{\mu_H}}\ge\int\widehat H\dd\sigma_{g_\mu}$,
and since both spectral measures have mass one and $\mu_H$ satisfies the
diagonal inequality,
$\int T\dd\sigma_{g_\mu}\ge\int T\dd\sigma_{g_{\mu_H}}-2\epsilon\ge-2\epsilon$.
Thus the inequality holds for the fixed $\mu$, although $\mu_H$ depends
on the packet. Cubic symmetry of $\mu$ gives the other five diagonal
normals.

Finally let $\delta\downarrow0$ along a convergent subsequence. The
$\ell^1$ change of coupling is $O(\delta)$, so the finite-window Gibbs
kernels converge uniformly in the exterior spins, the limit is a
nearest-neighbour Gibbs state at $\beta$, and it retains the symmetries and
all the finite pair inequalities. By Proposition~\ref{prop:boxes} its dot
correlation is a strongly reflection-positive kernel. \qed

\section{Lattice points for a prescribed free-cube limit}\label{app:freecube}

\begin{proposition}\label{prop:freecube-points}
In the situation of Corollary~\ref{cor:freecube}, \eqref{eq:pointwise}
holds for $g$, with its own susceptibility $\chi_R$, and so does its
annular version.
\end{proposition}

\begin{proof}
\emph{The even sublattice.} Put $G(z)=g(2z)$. Restricting the site forms
of $g$ to lists in $(2\Z)^3$ gives \ref{K4} for $G$; \ref{K1}--\ref{K3}
are inherited. For \ref{K5} let $\rho=\vt_{e_1}$; for $u_1,v_1\ge1$,
\[
 (2u-e_1)-\rho(2v-e_1)=2\bigl(u-\vtb v\bigr),
\]
and $2u-e_1\in\Hp_{e_1}$, so the axial site form of $g$ at the points
$2u-e_1$ is the bond form of $G$. Thus $G$ is strongly reflection
positive, and Theorem~\ref{thm:main}(ii) applies to $G$ with its susceptibility
$\chi^G_r$. With $r=\lfloor R/2\rfloor$ we have $\chi^G_r\le\chi_R$ and
\[
 \frac{R^3}{\chi_R}|g(2u)-g(2u')|=\Bigl(\frac Rr\Bigr)^3\frac{\chi^G_r}{\chi_R}
 \cdot\frac{r^3}{\chi^G_r}|G(u)-G(u')|,
\]
which proves \eqref{eq:pointwise} for $k_j,\ell_j\in(2\Z)^3$.

\emph{Profiles on the eight classes.} If $|x_i|=\|x\|_\infty=r$, the site
reflection in $x_i=\lfloor r/2\rfloor$ and Cauchy--Schwarz give
$g(x)^2\le G(\lfloor r/2\rfloor e_i)G(\lceil r/2\rceil e_i)$, so
Lemma~\ref{lem:pointbound} for $G$ yields $U_g(r)\le C\chi_r/r^3$. By
\ref{O2} for $g$ (Proposition~\ref{prop:admissible}) and the path
argument with even steps, the functions $R^3g/\chi_R$ are equicontinuous
on each class $p\in\Z^3/2\Z^3$ on compact annuli, so along a common
subsequence they converge to continuous profiles $f_p$; the even profile is
radial by the first step.

\emph{The infrared bound.} We may take $\beta>0$. By compactness realize
$g$ as the two-point function of a zero-field Gibbs law, and let $\sigma$ be
its spectral measure, $\dd k_*$ normalized Haar measure on $\mathbb T^3$
and $\lambda(k)=2\sum_i(1-\cos k_i)$. We claim
$\dd\sigma(k)\le\frac N{\beta\lambda(k)}\dd k_*$ off $k=0$. For real
finitely supported $b$ and $A=-\Delta$ choose $m\asymp L^3$ translates
$b_z$ in an even periodic cube of side $L$ with disjoint
$1$-neighbourhoods, a unit internal direction $u$, and
$X_z=\sum_x(Ab_z)(x)\,u\cdot S_x$. Gaussian domination on the torus
\cite[Thm.~2.1]{FSS} bounds $\E_{\rm per}\exp(t\sum_z\varepsilon_zX_z)$ by
$\exp(t^2m\ip b{Ab}/(2\beta))$ for all signs. For every exterior
configuration the cube Gibbs density relative to the periodic law is at
most $e^{18\beta L^2}$, by counting crossing and wraparound bonds.
Averaging the signs, Jensen's inequality bounds
$\sum_z\E\log\cosh(tX_z)$ by $18\beta L^2+t^2m\ip b{Ab}/(2\beta)$;
summing over an internal orthonormal basis, letting $L\to\infty$ and
then $t\to0$ gives
$\int\lambda^2|\widehat b|^2\dd\sigma\le(N/\beta)\int\lambda|\widehat b|^2\dd k_*$,
and polarization and approximation give the claim.

\emph{Equality of the profiles.} Let $\phi$ be smooth with compact support
in $\R^3\setminus\{0\}$ and $c\in\Z^3/2\Z^3$, $c\ne0$. Then
$\sum_x(-1)^{c\cdot x}g(x)\phi(x/R)=O_\phi(1)$: by Poisson summation the
Fourier series of the test is bounded in $L^1(\dd k_*)$ near $\pi c$ and
is $O_{\phi,M}(R^{3-M})$ away from it, while the spectral density is
bounded near $\pi c\ne0$. If $\chi_R\to\infty$, dividing by $\chi_R$ gives
zero in the limit; if $\chi_R$ stays bounded, the same follows from the
summable tail of $g\ge0$ on the support of the rescaled test. Riemann sums
give $\sum_p(-1)^{c\cdot p}\int\phi f_p=0$ for all $c\ne0$, and Fourier
inversion on the eight classes shows that all $f_p$ equal the even,
radial profile. The subsequence argument of Corollary~\ref{cor:points}
completes the proof.
\end{proof}

\section{Formal verification}\label{app:formal}

The main results of this paper have been verified in the Lean~4 proof
assistant with the Mathlib library. The formal statements are the
two parts of the abstract Theorem~\ref{thm:main}, with exactly the
hypotheses \ref{K1}--\ref{K4}, respectively \ref{K1}--\ref{K5}, and their
applications in Theorems~\ref{thm:ising}, \ref{thm:potts} and
\ref{thm:sphere}(i), (iii), (iv). For Theorem~\ref{thm:sphere}(ii) the formal
statement concerns every accumulation point of $P^e_{\beta,h}$ as
$h\downarrow0$; the identification of this limit in
Proposition~\ref{prop:heis-ident} is proved in the text.

In every formal statement the tangent is an arbitrary locally finite
vague limit of the measures \eqref{eq:muR} along the given sequence,
normalized by the susceptibility of the same kernel; $O$ ranges over all
linear isometries of Euclidean $\R^3$; and the atom at the origin is
retained. The code is available at \url{https://github.com/whyihaveyou/rp-rotation-lean}
and archived at Zenodo, version~1.0.0, \href{https://doi.org/10.5281/zenodo.24557203}{doi:10.5281/zenodo.24557203}.
It uses Lean \texttt{v4.34.0-rc2} and Mathlib at the same tag, and consists
of $188$ modules. After \texttt{lake exe cache get} and
\texttt{lake build}, the script \nolinkurl{scripts/check_paper_statements.py}
elaborates every declaration in Table~\ref{tab:lean} and prints its
axioms, and \nolinkurl{scripts/check_model_endpoints.py} checks $50$
statements specialized to the physical models and $105$ further named
results. All of them depend only on the standard axioms \texttt{propext},
\texttt{Classical.choice} and \texttt{Quot.sound}. All declarations lie in
the namespace \texttt{\small Ising\allowbreak{}Scaling}.

\begin{table}[p]
\centering\scriptsize
\renewcommand{\arraystretch}{1.2}
\begin{tabularx}{\textwidth}{@{}>{\raggedright\arraybackslash}p{3.1cm}
 >{\raggedright\arraybackslash}X >{\raggedright\arraybackslash}p{3.6cm}@{}}
\toprule
Statement & Lean declaration & Module\\
\midrule
Theorem~\ref{thm:main}(i) & \texttt{\small Model\allowbreak{}Packets.\allowbreak{}isotropic\_\allowbreak{}of\_\allowbreak{}site\_\allowbreak{}forms},
 \texttt{\small Model\allowbreak{}Packets.\allowbreak{}exists\_\allowbreak{}isotropic\_\allowbreak{}of\_\allowbreak{}site\_\allowbreak{}forms} & \texttt{\small Model\allowbreak{}Packet\allowbreak{}Physical\allowbreak{}Rotation}\\
Theorem~\ref{thm:main}(ii) & \texttt{\small Abstract\allowbreak{}Kernel.\allowbreak{}raw\_\allowbreak{}angular} & \texttt{\small Abstract\allowbreak{}Kernel\allowbreak{}Pointwise}$^*$\\
Corollary~\ref{cor:moments} & \texttt{\small Packet.\allowbreak{}grid\allowbreak{}Packet\_\allowbreak{}spectrum\_\allowbreak{}of\_\allowbreak{}site\_\allowbreak{}reflection} & \texttt{\small Reflected\allowbreak{}Lattice\allowbreak{}Packet}\\
Prop.~\ref{prop:admissible}, \ref{O1} & \texttt{\small Raw\allowbreak{}Lattice\allowbreak{}Oscillation.\allowbreak{}local\allowbreak{}Oscillation\_\allowbreak{}of\_\allowbreak{}site\_\allowbreak{}bond} & \texttt{\small Raw\allowbreak{}Lattice\allowbreak{}Oscillation}\\
Section~\ref{sec:site} & \texttt{\small Parity\allowbreak{}Smoothing.\allowbreak{}normalized\allowbreak{}Packet\allowbreak{}Source}, \texttt{\small local\allowbreak{}Oscillation\_\allowbreak{}of\_\allowbreak{}site\_\allowbreak{}forms} & \texttt{\small Rotor\allowbreak{}Packet\allowbreak{}Parity\allowbreak{}Smoothing}, \texttt{\small Model\allowbreak{}Packet\allowbreak{}Smoothed\allowbreak{}Oscillation}\\
Cor.~\ref{cor:points} & \texttt{\small Raw\allowbreak{}Lattice\allowbreak{}Oscillation.\allowbreak{}angular\_\allowbreak{}of\_\allowbreak{}local\allowbreak{}Oscillation},
 \texttt{\small annular\_\allowbreak{}of\_\allowbreak{}local\allowbreak{}Oscillation} & \texttt{\small Raw\allowbreak{}Lattice\allowbreak{}Angular\allowbreak{}Readout}\\
Prop.~\ref{prop:boxes} (limits) & \texttt{\small Packet.\allowbreak{}site\allowbreak{}Reflected\allowbreak{}Positive\_\allowbreak{}of\_\allowbreak{}tendsto} & \texttt{\small Reflected\allowbreak{}Lattice\allowbreak{}Packet}\\
\midrule
Thm.~\ref{thm:potts}, Potts & \texttt{\small Potts.\allowbreak{}Site\allowbreak{}Box.\allowbreak{}canonical\allowbreak{}Free\allowbreak{}Pair\_\allowbreak{}isotropic},
 \texttt{\small canonical\allowbreak{}Free\allowbreak{}Pair\_\allowbreak{}raw\_\allowbreak{}angular}, \texttt{\small canonical\allowbreak{}Free\allowbreak{}Pair\_\allowbreak{}axial\allowbreak{}Bond\_\allowbreak{}positive}
 & \texttt{\small Model\allowbreak{}Packet\allowbreak{}Physical\allowbreak{}Rotation}, \texttt{\small Raw\allowbreak{}Lattice\allowbreak{}Angular\allowbreak{}Readout}, \texttt{\small Potts\allowbreak{}Canonical\allowbreak{}Shifted\allowbreak{}Mirror}\\
Thm.~\ref{thm:potts}, FK & \texttt{\small free\allowbreak{}Law\_\allowbreak{}tangent}, \texttt{\small exhaustion\_\allowbreak{}tendsto\_\allowbreak{}free\allowbreak{}Law},
 \texttt{\small Edwards\allowbreak{}Sokal.\allowbreak{}free\_\allowbreak{}connectivity\_\allowbreak{}mono} & \texttt{\small Potts\allowbreak{}Free\allowbreak{}Bond\allowbreak{}State}, \texttt{\small Potts\allowbreak{}Free\allowbreak{}Bond\allowbreak{}Exhaustion}, \texttt{\small Potts\allowbreak{}Free\allowbreak{}FKLimit}\\
Thm.~\ref{thm:ising} & \texttt{\small Abstract\allowbreak{}Kernel\allowbreak{}Instance.\allowbreak{}ising\_\allowbreak{}isotropic\_\allowbreak{}via\_\allowbreak{}kernel};
 independently \texttt{\small Is\allowbreak{}Two\allowbreak{}Point\allowbreak{}Tangent.\allowbreak{}rotation\_\allowbreak{}invariant} (MMS route)
 & \texttt{\small Ising\allowbreak{}Abstract\allowbreak{}Kernel\allowbreak{}Instance}$^*$, \texttt{\small Ising\allowbreak{}Two\allowbreak{}Point\allowbreak{}Rotation}\\
Thm.~\ref{thm:sphere}(i) & \texttt{\small Rotor.\allowbreak{}XY.\allowbreak{}graph\allowbreak{}Pair\_\allowbreak{}mono}, \texttt{\small canonical\allowbreak{}Free\allowbreak{}Gibbs\_\allowbreak{}raw\_\allowbreak{}angular}
 & \texttt{\small Rotor\allowbreak{}XYGinibre}, \texttt{\small Rotor\allowbreak{}XYFree\allowbreak{}Gibbs\allowbreak{}Rotation}\\
Thm.~\ref{thm:sphere}(ii)$^\dagger$ & \texttt{\small Heisenberg\allowbreak{}Ferromagnetic\allowbreak{}Lee\allowbreak{}Yang.\allowbreak{}partition\_\allowbreak{}ne\_\allowbreak{}zero},
 \texttt{\small Physical\allowbreak{}Rotation.\allowbreak{}exists\_\allowbreak{}vanishing\_\allowbreak{}limit}, \texttt{\small sharp\_\allowbreak{}rotation}, \texttt{\small rounded\_\allowbreak{}rotation}
 & \texttt{\small Heisenberg\allowbreak{}Ferromagnetic\allowbreak{}Lee\allowbreak{}Yang}, \texttt{\small Rotor\allowbreak{}Heisenberg\allowbreak{}Physical\allowbreak{}Rotation}\\
Thm.~\ref{thm:sphere}(iii) & \texttt{\small Rotor.\allowbreak{}Pressure\allowbreak{}Selection.\allowbreak{}ONFamily.\allowbreak{}exists\_\allowbreak{}on\_\allowbreak{}family\_\allowbreak{}rotation\_\allowbreak{}full} & \texttt{\small Rotor\allowbreak{}ONAll\allowbreak{}NRotation}\\
Thm.~\ref{thm:sphere}(iv) & \texttt{\small exists\_\allowbreak{}unique\_\allowbreak{}nearest\_\allowbreak{}gibbs\_\allowbreak{}tanh} & \texttt{\small Rotor\allowbreak{}High\allowbreak{}Temperature\allowbreak{}Uniqueness}, \texttt{\small Rotor\allowbreak{}Free\allowbreak{}Unique\allowbreak{}Exhaustion}\\
Cor.~\ref{cor:freecube}, App.~\ref{app:freecube}$^\ddagger$ & \texttt{\small Is\allowbreak{}Stationary\allowbreak{}Free\allowbreak{}Cube\allowbreak{}Pair\allowbreak{}Limit.\allowbreak{}surface\_\allowbreak{}isotropic},
 \texttt{\small raw\_\allowbreak{}angular\_\allowbreak{}even\_\allowbreak{}lattice} & \texttt{\small Model\allowbreak{}Packet\allowbreak{}Physical\allowbreak{}Rotation}, \texttt{\small Rotor\allowbreak{}Stationary\allowbreak{}Raw\allowbreak{}Angular}\\
Prop.~\ref{prop:ising-relative} & relative comparison & \texttt{\small Ising\allowbreak{}Relative\allowbreak{}Angular\allowbreak{}Oscillation}\\
\bottomrule
\end{tabularx}
\caption{Statements and formal declarations. $^*$Files composing lemmas of
the package.
$^\dagger$Lean proves the statement for every vanishing-field accumulation
state; Proposition~\ref{prop:heis-ident} (existence of the full limit and
identification with the fixed-boundary state) is not formalized, nor is
the geometric argument giving exhaustion independence of the free Ising
limit for arbitrary domains. $^\ddagger$Lean proves the lattice-point
comparison on $(2\Z)^3$, and on all of $\Z^3$ under the additional
hypothesis that the bond-centred rectangles have the same limit; the
eight-class argument of Appendix~\ref{app:freecube} is not formalized.}
\label{tab:lean}
\end{table}

\end{document}